\documentclass[12pt,a4paper]{article}
\usepackage{hyperref}
\usepackage{graphicx}
\usepackage{subfigure}
\usepackage{fancyhdr}
\usepackage{natbib}
\begin{document}

\section* {Magnetic reconnection at 3D null points: effect of magnetic field asymmetry}\ \ \ \ \ \ \ \ \ \ \ \ \ \ \ \ \ \ \ \ \ \ \ 
A. K. Al-Hachami and D. I. Pontin\\
Division of Mathematics, University of Dundee, U.K.

\section*{Abstract}
The magnetic field in many astrophysical plasmas, for example in the solar corona, is known to have a highly complex -- and clearly three-dimensional -- structure. Turbulent plasma motions in high-$\beta$ regions where field lines are anchored, such as the solar interior, can store large amounts of energy in the magnetic field. This energy can only be released when magnetic reconnection occurs. Reconnection may only occur in locations where huge gradients of the magnetic field develop, and one candidate for such locations are magnetic null points, known to be abundant for example in the solar atmosphere. Reconnection leads to changes in the topology of the magnetic field, and energy released as heat, kinetic energy and acceleration of particles. Thus reconnection is responsible for many dynamic processes, for instance flares and jets. 
The aim of this paper is to investigate the properties of magnetic reconnection at a 3D null point, with respect to their dependence on the symmetry of the magnetic field around the null. In particular we examine the rate of flux transport across the null point with symmetric/asymmetric diffusion regions, as well as how the current sheet forms in time, and its properties.
Mathematical modelling and finite difference  resistive MHD simulations. 
It is found that the basic structure of the mode of magnetic reconnection considered is unaffected by varying the magnetic field symmetry, that is, the plasma flow is found  cross both the spine and fan of the null. However, the peak intensity and dimensions of the current sheet are dependent on the symmetry/ asymmetry of the field lines. As a result, the reconnection rate is also found to be strongly dependent on the field asymmetry. 
The symmetry/asymmetry of the magnetic field in the vicinity of a magnetic null can have a profound effect on the geometry of any associated reconnection region, and the rate at which the reconnection process proceeds.
\section{Intoduction} 
Magnetic reconnection is the breaking and topological or geometrical rearrangement of the magnetic field lines in a plasma. The magnetic field plays a fundamental  role in many of the phenomena that occur in the plasma. It is not surprising that three-dimensional (3D) magnetic fields are more complex than two-dimensional ones. 
It is known from observations that magnetic reconnection occurs in abundance in astrophysical plasmas. However, due to the very low plasma resistivity, reconnection may only occur where very intense currents (`current sheets') develop. One of the most fundamental questions that must be answered to determine the locations and mechanisms of energy release in astrophysical plasmas is therefore: where may such  currents develop? 

It is now becoming clear that the magnetic field in the solar atmosphere has a highly complex structure. Two major candidates that have been proposed as sites of current sheet formation in such a complex magnetic field are 3D nulls points, and associated separator field lines \citep{lau1990, klapper1996, priest1996, longcopecowley1996, galsgaard1997, pontincraig2005,longcope1996,longcope2001} -- field lines that link two nulls. We focus here on reconnection at isolated 3D nulls.
Indications are that an abundance of 3D nulls is present in the solar corona \citep{regnier2008, longcope2009}, which have been suggested as likely sites for coronal heating \cite[e.g.][]{priest2005}. Moreover, recent observations suggest that reconnection at such nulls may play an important role in jets \citep{pariat2009,torok2009} solar flares \citep[e.g.][]{luoni2007,masson2009} and coronal mass ejections \citep[e.g.][]{ugarteurra2007, barnes2007}. Furthermore, recently the first in-situ observations have been made by the Cluster spacecraft of single and multiple 3D magnetic nulls in the Earth's magnetotail \citep[e.g.][]{xiao2006}. The  observations further suggest that these nulls may be playing an important role in the reconnection process occurring in the magnetotail. 
Though magnetic field measurements in other astrophysical objects further afield are difficult, it is almost certain that similar reconnection processes at 3D nulls also occur there.

 To find the local magnetic structure about a null point, we consider the magnetic field in the vicinity of null point where the field vanishes ($\bf{B}={\bf 0}$). If  the null point is taken to be situated at the origin and, in addition, we assume we are sufficiently close to the null, then the magnetic field may be expressed as 
      \begin{equation} 
      {\bf{B}}=\mathcal{M}\cdot \bf{r} \label{equ1}
      \end{equation}
 where $\mathcal{M}$ is a matrix with the elements of  the Jacobian of {\bf{B}} and $\bf{r}$ is the position vector $(x,y,z)^T$. The eigenvalues of $\mathcal{M}$ sum to zero since $\nabla \cdot {\bf B}=0$. We  consider the situation where all the eigenvalues are real. Since they sum to zero there is always one eigenvalue of opposite sign to the other two. The two eigenvectors corresponding to the eigenvalues with same-sign real part define the ``fan surface" of the null. The third eigenvector defines the orientation of the ``spine line". For more details, see e.g.~\cite{fukao1975, lau1990, parnell1996}.

Unlike in two dimensions, reconnection can occur in 3D either at a null point or in the absence of nulls \citep{schindler1988, priest2000,Demoulin2006}. What's more, the nature of reconnection in 3D has been shown to be fundamentally different from 2D reconnection \citep{priesthornig2003}. The nature of magnetic reconnection in the absence of a three-dimensional null point has been discussed by \cite{hesse1991} and \citet{Hornig2003}. 
The kinematics of steady reconnection at three dimensional null points have been studied by \cite{priest1996} when  $\eta=0$. Later, \cite{spine2004, fan2004} improved this model by adding a finite  resistivity, localised around the null point. Two distinct cases were considered, in which the current ($\bf{J}$) was directed parallel to first the spine and second the fan plane of the null.
The structures of the two solutions were found to differ greatly, and as a result, the reconnection rate, calculated by integrating the ${E_{||}}$ along field lines, represents very different behaviors of the flux for the two cases. 
In the first case, in which $\bf{J}$ was directed parallel to the spine, a type of rotational flux mis-matching was found, with no flow being present across either the spine or the fan of the null point. On the other hand, when $\bf{J}$ was directed parallel to the fan surface, it was found that magnetic flux is transported through the spine line and the fan plane, in a process much more conceptually similar to the 2D case. 
In this case it can be shown that the reconnection rate gives a measure of the rate of flux transport across the separatrix surface of the null \citep{fan2004}. The case in which $\bf{J}$ is parallel to the spine corresponds to one pair of complex conjugate eignevalues, whereas when ${\bf J}$ is parallel to the fan the eignevalues are all real.
In each of these investigations only the azimuthually symmetric case was considered, that is the case in which the magnetic field in the fan plane is isotropic.    
In this paper we focus on the case where ${\bf J}$ is parallel to the fan surface (real eigenvalues), and for the first time consider magnetic reconnection at a generic non-symmetric magnetic null point, i.e.~a null for which the fan eigenvalues are not equal. The different modes of reconnection that occur in practice in a plasma (when the full set of MHD equations are considered) have recently been classified by \cite{priest2009}. In terms of the framework they have set up, the mode of reconnection considered here is termed {\it spine-fan reconnection}. In a future paper we will go on to generalise the complex conjugate eigenvalue case.

In sections \ref{kinematic} and \ref{analsec}, we describe a kinematic model for reconnection at a non-symmetric null point,  comparing our results with those of \cite{fan2004}. In section \ref{numerical} we describe the results of a related resistive magnetohydrodynamic (MHD) numerical simulation,  and in section \ref{conc} we present our conclusions.
\section{Kinematic solution -- method}\label{kinematic}
\subsection{The model}
The subject of magnetic reconnection  is a complex one, and its study is  still in the early stages. Therefore, one approach that is used to try to understand the properties of this process is to consider  a reduced set of the MHD equations. There are a number of analytical 3D solutions, which are described by \cite{Hornig2003} and \cite{antonia2006,antonia2009}, where there is no null point of the magnetic field, as well as the solutions in the presence of a null mentioned above \citep{spine2004, fan2004, priest2009}. These solutions are kinematic reconnection, that is they satisfy  Maxwell's equations, as well as the induction equation. 
This approach can give great insight into the topological structure of a magnetic reconnection process occurring at an isolated diffusion region \citep{schindler1988}. After investigating the properties of the solutions of this subset of the MHD equations, we go on in section \ref{numerical} to examine which properties survive when the full set of resistive MHD equations is solved.
   
\begin{figure}
\begin{center}
\subfigure[]{\label{fig:edge-3b}\includegraphics[width=0.42\textwidth]{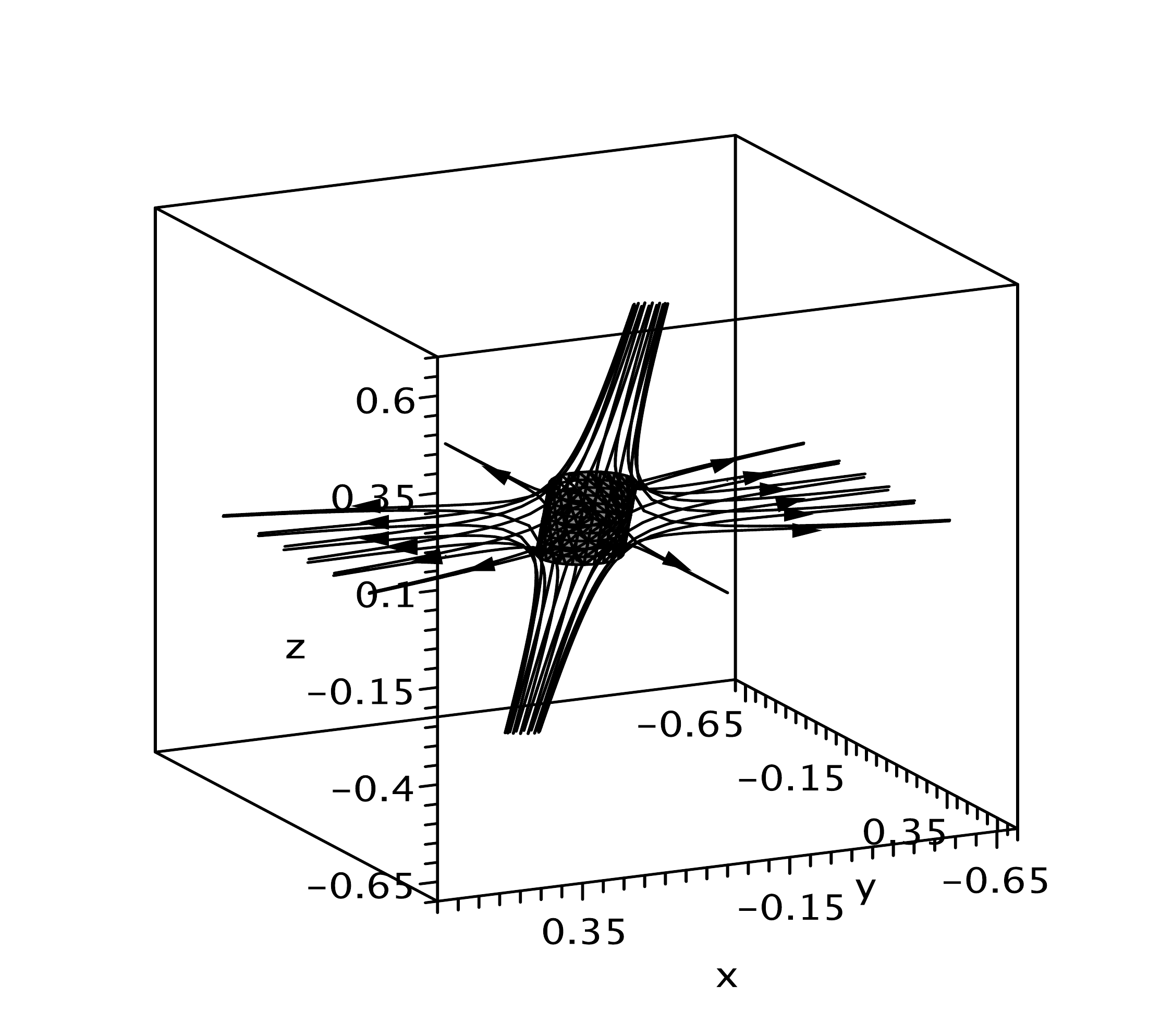}}
\subfigure[ ]{\label{fig:edge-3c}\includegraphics[width=0.42\textwidth]{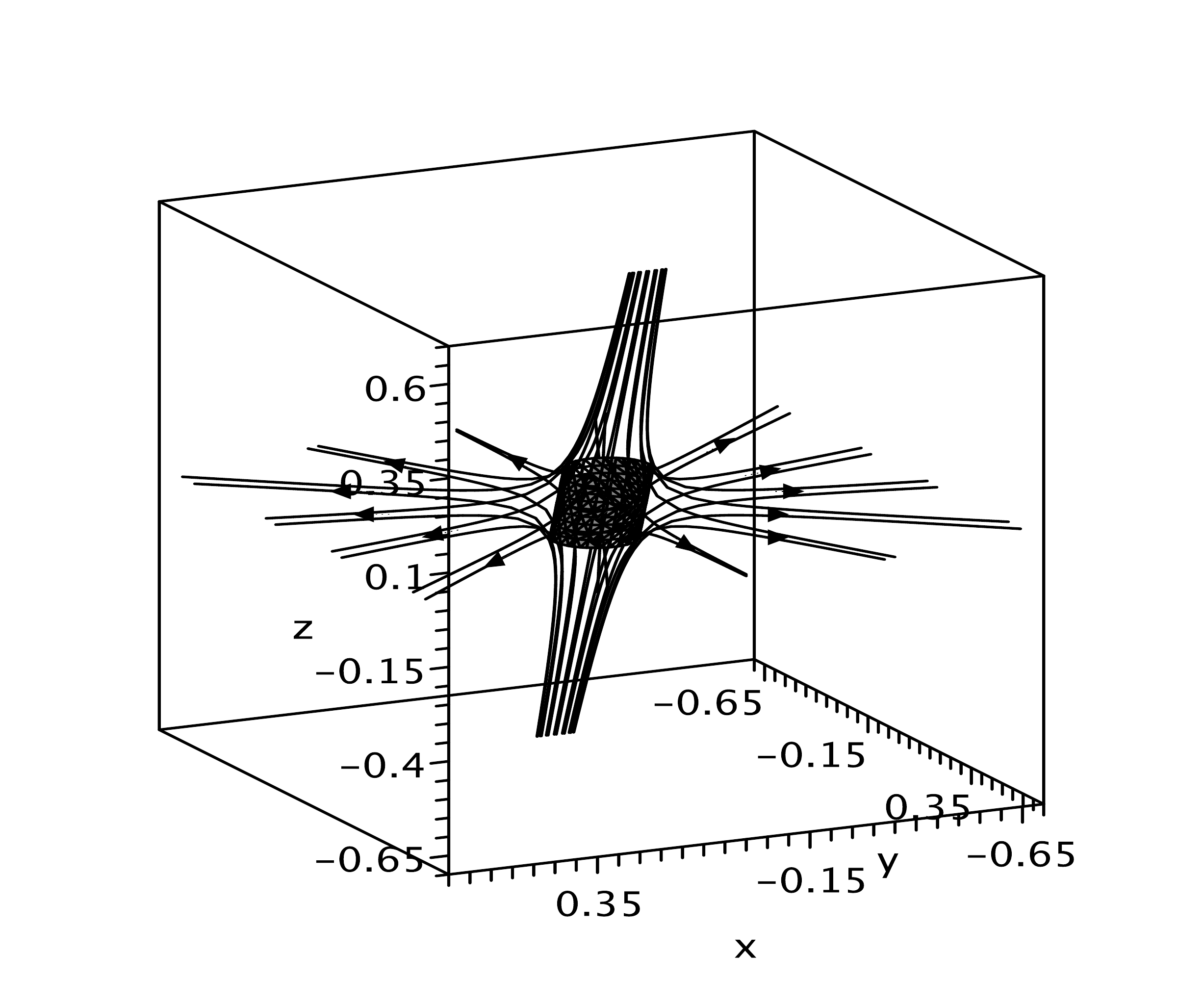}}
\subfigure[]{\label{fig:edge-3a}\includegraphics[width=0.4\textwidth]{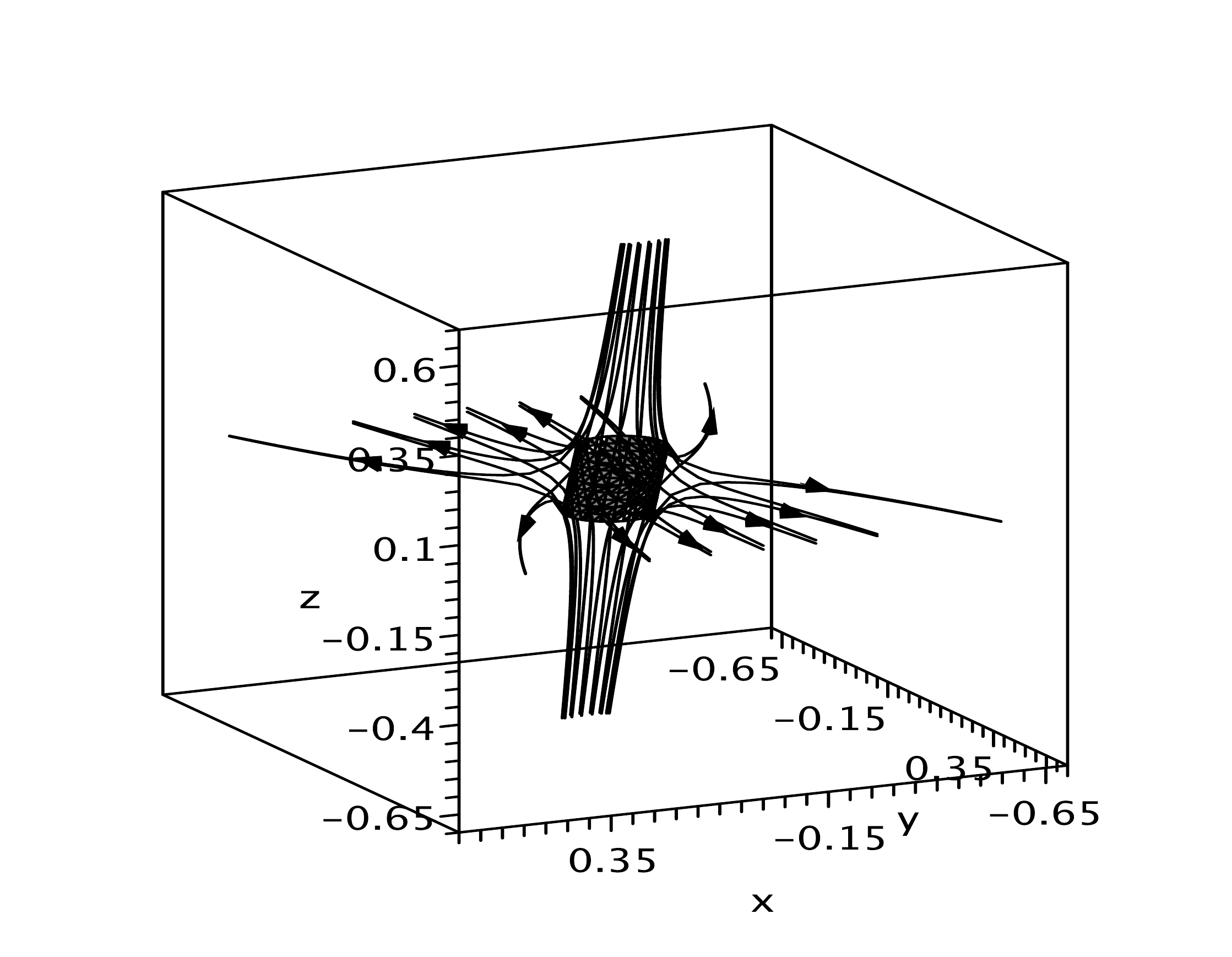}}
\end{center}
\caption{\footnotesize The structure of the magnetic null point with $j=1$ and different values of $p$: (a) $p=0.5$, (b) $p=1$ and (c) $p=2$. }
\label{ali1r}
  \end{figure}

 We seek a solution to the kinematic, steady, resistive MHD equations in the locality of a magnetic null point. That is, we solve 
\begin{eqnarray} 
      \bf{E+v\times B = \eta J}  \label{equ8}\\
\nabla \times \bf{E}=0  \label{equ4}\\
        \nabla \times \bf{B}=\mu _{0}{\bf J} \label{1}\\
\nabla\cdot \bf{B}=0 \label{x}
 \end{eqnarray}
As discussed above, here we consider a null point with current directed parallel to the fan plane. We choose the magnetic field to be
\begin {equation}
{\bf{B}}=\frac{B_{0}}{L}\frac {2}{p+1}(x,py - jz,- (p+1)z) \label{equ3}
\end{equation}
where $p$ is a parameter (here we restrict ourselves to the case $p>0$). This generalises the previous work by \cite{fan2004}, who considered only the case where the field in the fan plane ($z=0$) is azimuthally symmetric, corresponding to $p=1$. For convenience we will write $2B_{0}/L(p+1)=B_{0}^{\prime}$.  The current lies in the $x$-direction, and is given by $ {\bf{J}}=(B_{0}^{\prime}/\mu_{0})(j,0,0) $ from Eq.~(\ref{1}).  
Examining the matrix $\mathcal{M}$ (see Eq.~$\ref {equ1}$), the eigenvalues of the null point are found to be
\begin{displaymath} 
\lambda_{1}={B_{0}^{\prime}}, \qquad  \lambda_{2}=pB_{0}^{\prime}, \qquad \lambda_{3}=-(p+1)B_{0}^{\prime}
\end{displaymath} 
with corresponding eigenvectors 
\begin{displaymath} 
{\bf{k}}_1 =(1,0,0), \qquad
 {\bf{k}}_2 =(0,1,0), \qquad
  {\bf{k}}_3 =\left(0,1,\frac{2p+1}{j} \right)
\end{displaymath} 
It is clear from the above that the fan plane is defined by ${\bf k}_{1}$ and  ${\bf k}_{2}$  (since $p>0$). The fan plane of this magnetic null point is coincident with the plane $z=0$ while the spine is not perpendicular to this, but rather lies along $x=0, y=jz/(2p+1)$ (see figure $\ref {ali1r}$). 

For the chosen magnetic field (\ref{equ3}), closed-form expressions for the equations of
magnetic field lines can be found, by solving  
\begin{equation}
\frac{\partial {\bf{X}}(s) }{\partial s}={\bf{B(X}}(s)), \label{equ2}
\end{equation}  
where the parameter $s$ runs along field lines, to give
\begin{equation}\label{6a}\\ 
x =  x_{0}  {e ^{B_{0}^{\prime}s}} \end{equation}
\begin{equation}\label{6b}
y=\left (y_{0}-\frac{jz_{0}}{2p+1}\right){e^{pB_{0}^{\prime}s}}+\frac{jz_{0}}{2p+1}{e^{-B_{0}^{\prime}(p+1)s}}\end{equation}
\begin{equation}\label{6c}
z =  z_{0}  {e ^{-B_{0}^{\prime}(p+1)s}} .
\end{equation}
The inverse of equations  (\ref{6a},\ref{6b},\ref{6c})  are
\begin{equation}\label{10}
x_{0} =  x  {e ^{-B_{0}^{\prime}s}} 
\end{equation}
\begin{equation}\label{11}
y_0=\left (y-\frac{jz}{2p+1}\right){e^{-pB_{0}^{\prime}s}}+\frac{jz}{2p+1}{e^{B_{0}^{\prime}(p+1)s}}
\end{equation}
\begin{equation}\label{12}
z_{0} =  z {e ^{B_{0}^{\prime}(p+1)s}} 
\end{equation}
which describes the equations of the magnetic field lines in terms of some initial coordinates ${\bf{X}}_{0}=(x_{0}, y_{0},z_{0}) $.

We proceed to solve (\ref{equ8}-\ref{x}) as follows. From equation (\ref{equ4}) we can write, in general $\bf{E}=-\nabla \phi$  where $\phi$ is a scalar potential. Then the component of equation (\ref{equ8}) parallel to {\bf{B}}  is $-(\nabla \phi)_{||} = \eta {\bf{J_{||}}}$ 
and we can calculate $ \phi$ by integrating along magnetic field lines: 
 \begin {equation}\label{13}
 \phi=-\int \eta \,{\bf{ J\cdot B} } \, ds +\phi_{0}
  \end{equation}
  where  $\phi_{0}$  is a constant of integration. By substituting the equations (\ref{6a},\ref{6b},\ref{6c}) into the integrand of Eq.~(\ref{13}), we can perform this integration to obtain $\phi({\bf{X}}_{0},s)$. One this is done, we use equations (\ref{10},\ref{11},\ref{12}) to eliminate $s, x_0$ and $y_0$ to obtain $\phi({\bf{X}})$, treating $z_0$ as a constant (see below). 
The electric field can subsequently be found from
   \begin {equation}\label{equ9}
 \bf{E}=-\nabla \phi
  \end{equation}
and we then find the plasma velocity perpendicular to the magnetic field  $\bf{v_{\perp}}$, by taking the vector product of equation (\ref{equ8}) with {\bf{B}} to obtain  
\begin{equation}\label{22}
{\bf{v_{\perp}}}=  \frac{({\bf{E-\eta J)\times B}}}{B^2} 
\end{equation}

Now, in order to investigate the properties of magnetic reconnection in a fully 3D system, we impose a resistivity model which ensures that the diffusion region is spatially localised in 3D. This is also the case relevant to astrophysical plasmas, which are known to be effectively ideal except in very small regions where energy release occurs. The diffusion region is chosen to be localised around the null point, in line with the results of past work which has shown that shearing motions tend to focus current in the vicinity of the null \citep{rickard1996, pontingalsgaard2007, pontinbhat2007a}. Since the current is uniform in our simple model, 
we choose the resistivity to be localised, and take it to be of the form
   \begin{equation}\label{etaprof}
\eta=\eta_{0}\left \{
\begin{array}{cc}
\left(\frac{R1^2}{a^2}-1\right)^2   \left( \frac{(z^2)^{\frac{2p}{p+1}}}{b^2}-1\right)^2 & ~~~~~ R^2<a^2, \, \, \,  ( z^2)^{\frac{2p}{p+1}}<b^2 \\
\left(\frac{R2^{2}}{a^{2/p}}-1\right)^2   \left( \frac{(z^2)^{\frac{2}{p+1}}}{b^2}-1\right)^2 & ~~~~~ R^2<a^2, \, \, \,  ( z^2)^{\frac{2}{p+1}}<b^2 \\

0 & \mbox{otherwise}
\end{array} \right.
\end{equation}
 where $R_1=\sqrt{({x}^2)^p+(y-jz/(2p+1))^2}$ and $R_2=\sqrt{{x}^2+((y-jz/(2p+1))^2)^{1/p}}$, where $\eta_{0}$, a and b are positive constants.
This is done in order to localise the product $\eta {\bf J}$, and hence the diffusion region, since we have not yet discovered a way to proceed with our analytical method with localised ${\bf J}$. The exact mathematical form for $\eta$ is not expected to affect the qualitative structure of the solution, and is chosen in order to render the equations tractable. The crucial property for the structure of the solution is the localisation of the diffusive term $\eta {\bf J}$.  The dependence of $\eta$ on $p$ is chosen differently for $p\geq 1$ and $p<1$ to ensure that $\eta(x,y,z)$ is always differentiable, and is chosen in such a way as to maintain consistency in the dependence of the diffusion region size in the $x$-direction on $p$, which is shown later to be an important property.
 $\eta_{0}$ is the value of $\eta$  at the null point, and the diffusion region is a tilted cylinder centered on the spine axis, extending to $z=\pm b^{(p+1)/2p}$  when $p\ge 1$ and $z=\pm b^{(p+1)/2}$ when $p< 1$. The cross-section of the diffusion region in the $z=0$ plane is circular with radius $a$ when $p=1$, but when $p \neq 1$, it extends to $x=\pm a^{1/p}$ and $y=\pm  a$. 
In order to integrate Eq.~(\ref{13}), we must choose a surface on which to start our integration (i.e.~on which to set $s=0$) that intersects each field line once and only once, in order that $\phi$ is single-valued. We choose surfaces above and below the fan surface, $z=\pm z_0$, constant. To simplify the mathematical expressions, and without loss of generality, we assume $z_0=b$. Performing the calculation of $\phi({\bf X})$ as described above yields two expressions for $\phi$,  for  $z>0$ and   $z<0 $. In order  to match these two expressions at the fan plane, that is for $\phi$ to be  smooth and continuous, and thus physically acceptable, we must set the value of $\phi$ at $z=\pm z_{0}$ (i.e. $\phi_{0}$ in Eq.~\ref{13}) to be 
{\begin{eqnarray}
 \phi_{0}=\phi_1 x_0 \left\{
        \begin{array}{lc}
\frac{1}{7}b^\frac{4(p-1)}{p+1} - \frac{2}{3}b^\frac{2(p-1)}{p+1}-1,&  0<p< 1,\\    
\frac{1}{(8p-1)}b^\frac{4(p-1)}{p+1} -\frac{2}{(4p-1)}b^\frac{2(p-1)}{p+1}-1,&  p\ge 1,      
        \end{array}\right.
\end{eqnarray}}
where
\begin{equation}\label{phi1}
\phi_1=  \frac{2B_{0}}{L(p+1)} \frac{\eta_{0} j }{\mu_0}
\end{equation}
$\phi$(X), $\bf{E}$ and $\bf{v_{\perp}} $ can be obtained from (\ref{13}),(\ref{equ9}) and (\ref{22}), as described earlier. The mathematical expressions are too lengthy to show here but can be calculated  using a symbolic computation package. Here we have used Maple v.12.

\section{Kinematic solution -- analysis}\label{analsec}
\subsection{Nature of reconnection}
\begin{figure}
\begin{center}
\subfigure[]{\label{fig:edge-3aa}\includegraphics[width=0.31\textwidth]{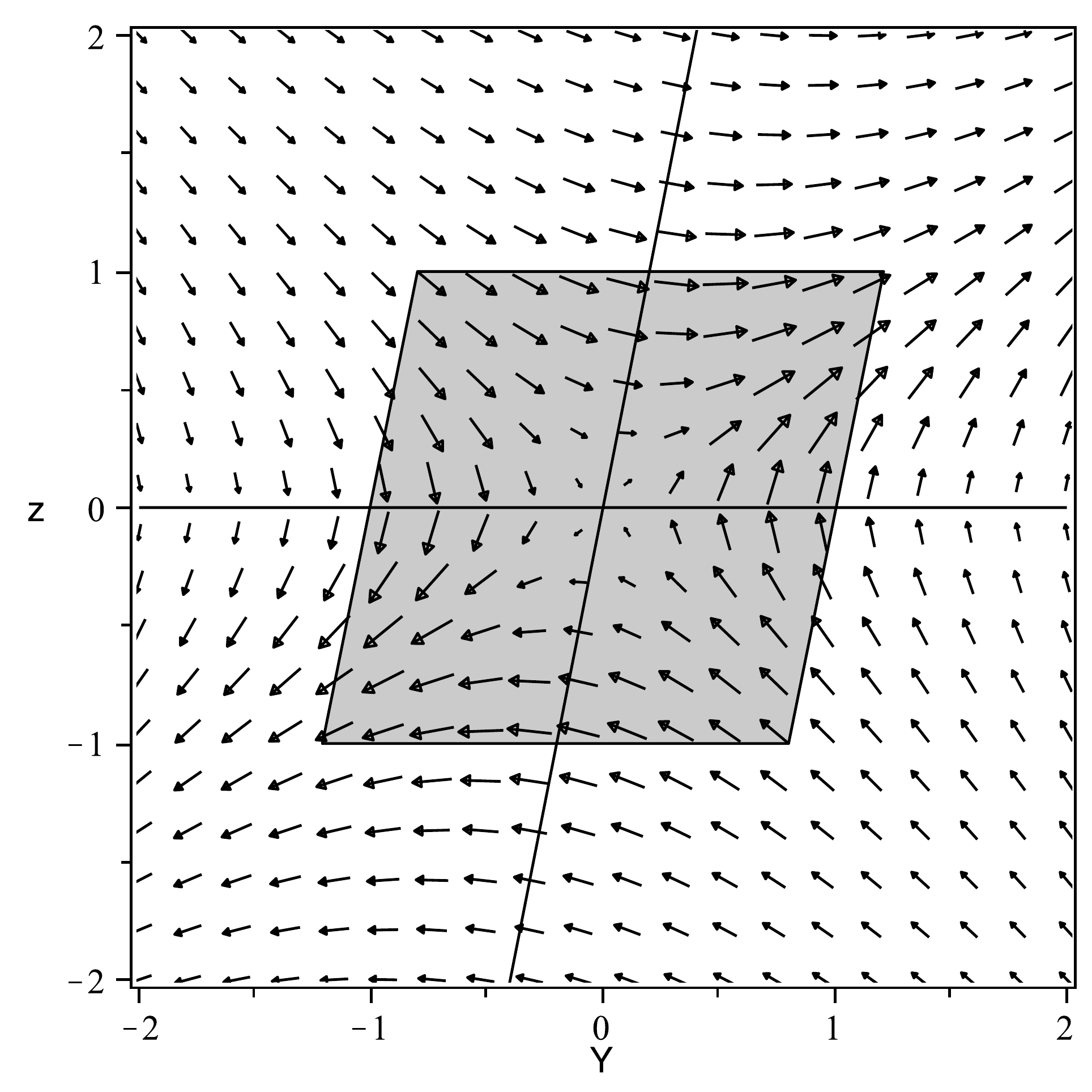}}
\subfigure[ ]{\label{fig:edge-3b}\includegraphics[width=0.31\textwidth]{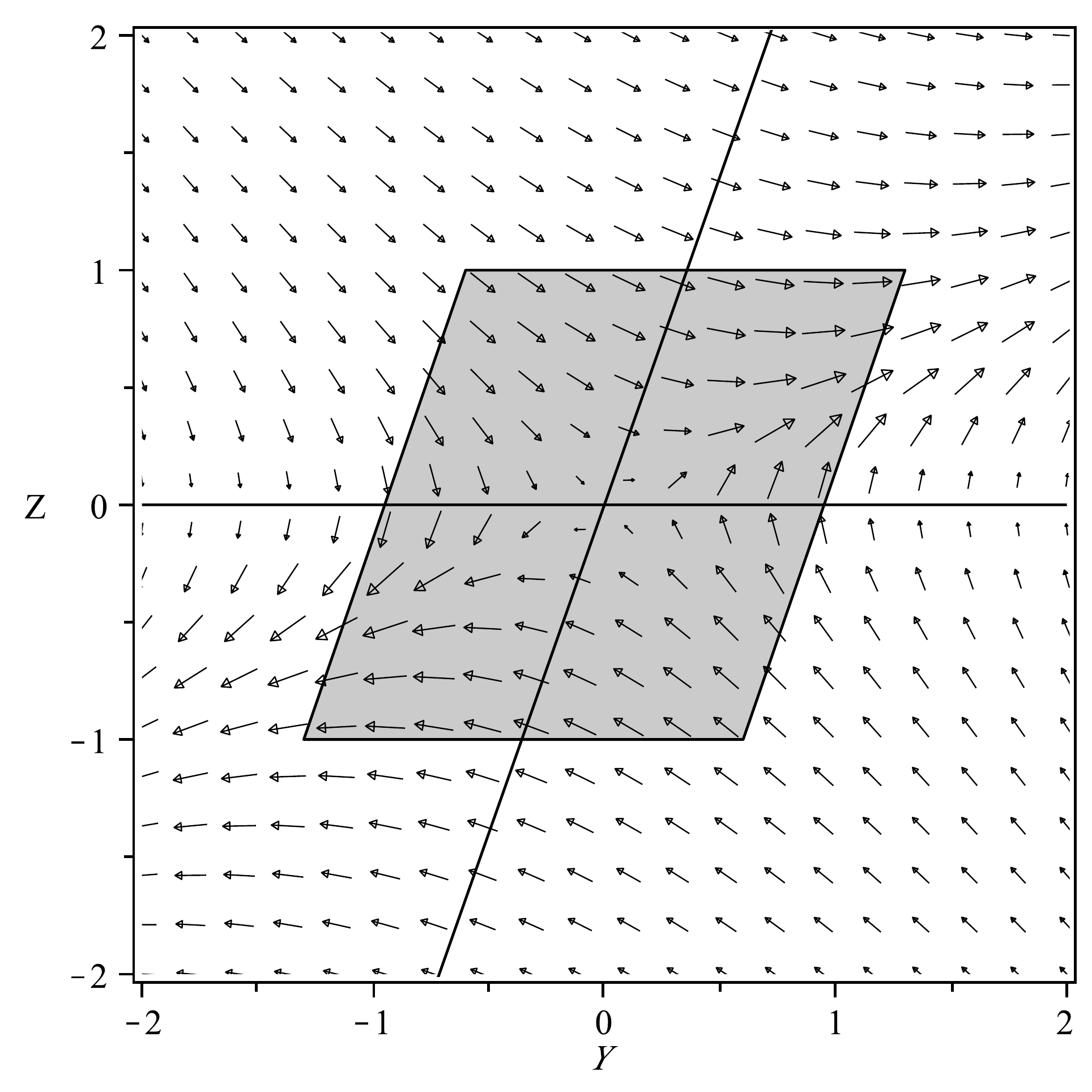}}
\subfigure[]{\label{fig:edge-3d}\includegraphics[width=0.31\textwidth]{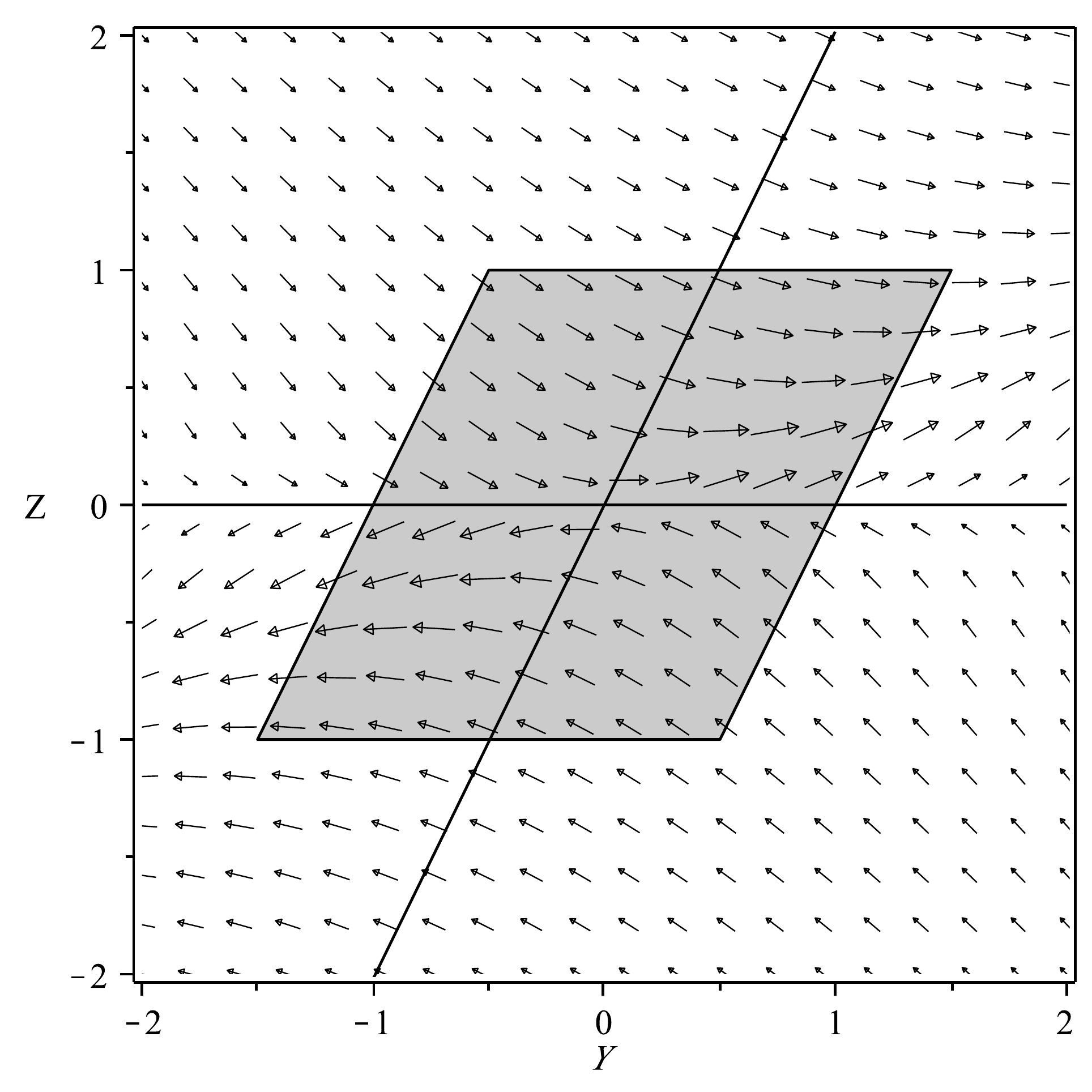}}
\end{center}
\caption{\footnotesize{Structure of the plasma flow across the spine and fan (black lines) in typical plane of constant $x=0$, {where the grayed area is the diffusion region}, for (a) $p=2$, (b) $p=0.9$, (c)  $p=0.5$, 
for parameters $\eta_{0}=\mu_0=B_0=j=a=b=L=1$.}} 
\label{ali2}
\end{figure}

In order to determine the structure of the magnetic reconnection process, we will examine the plasma velocity perpendicular to the magnetic field ($\bf{v_{\perp}}$). This velocity transports the magnetic flux outside the diffusion region. The flow does not cross the spine in the $x$-direction ($v_{\perp x}(0,y,z)=0$), so that $v_{\perp x}$ is negligible for the reconnection process. However, in the $yz$-plane, the plasma flow  crosses both the spine and the fan. Note that this is qualitatively the same as the situation described by \cite{fan2004} in the case of $p=1$. The nature of the plasma flow in a plane of constant $x$ with different values of $p$ is shown in Fig.~\ref{ali2}. Note that the qualitative structure -- of a stagnation-point flow -- is not affected a great deal by varying $p$. However the general trend is that as $p$ tends to zero, the plasma flow across the fan plane  becomes weaker. We will return to discuss this behaviour below.
\subsection {Reconnection rate}
\begin{figure} 
   \centering
   \includegraphics[width=8 cm]{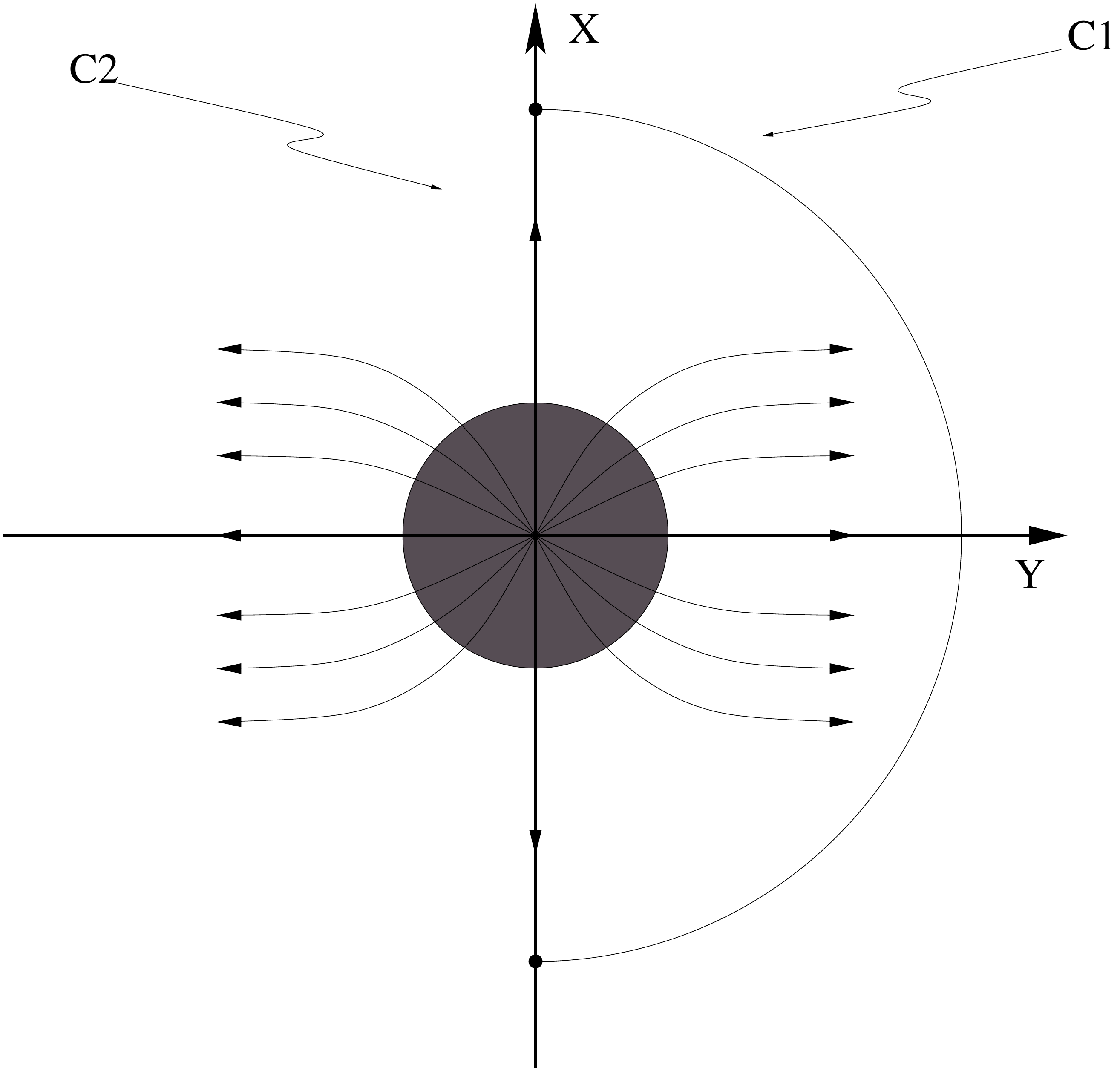} 
\caption{\footnotesize The curves C1 and C2 joining two points on the $x$-axis, {where the grayed area is the diffusion region, the arrows indicate the direction of field lines, for $a$=1 }    }
    \label{ali5}
\end{figure}

It is generally accepted that magnetic reconnection plays a fundamental role in many types of explosive astrophysical phenomena, for example solar flares. Yet what determines the reconnection rate is still a major problem and this is an important aspect of any reconnection model. In general, the reconnection rate in 3D is defined by the maximal value of 
\begin {equation}\label{recrateeq}
F=\int E_{\parallel} \, ds
\end{equation} 
along any field line threading a spatially localised diffusion region $D$ \citep[e.g.][]{schindler1988}. By symmetry, in this case 
\begin {equation}\label{a}
F=\int_{C2} E_{\parallel} \, ds
\end{equation} 
where the curve C2 lies along the $x$-axis, as shown in figure \ref{ali5}. Since the fan is a flux surface, the integral may equally well be performed along the curve C1, the curve C1 lying in the fan perpendicular to ${\bf B}$, see Fig.~\ref{ali5}. Now, since the curve C1 lies outside $D$ and therefore, along it $\bf{v}\times\bf{ B}= -\bf{E}$,  we can write
\begin{equation}\label{}
F= - \int_{C1} \bf{v}\times \bf{B}\cdot dl 
\end{equation}
from which it is clear that this reconnection rate measures the rate at which flux is transported across the fan surface by the flow in the ideal region  \citep{fan2004}.

\begin{figure}  
   	\centering
	\subfigure[]{\label{rr-a} \includegraphics[width=7.5 cm]{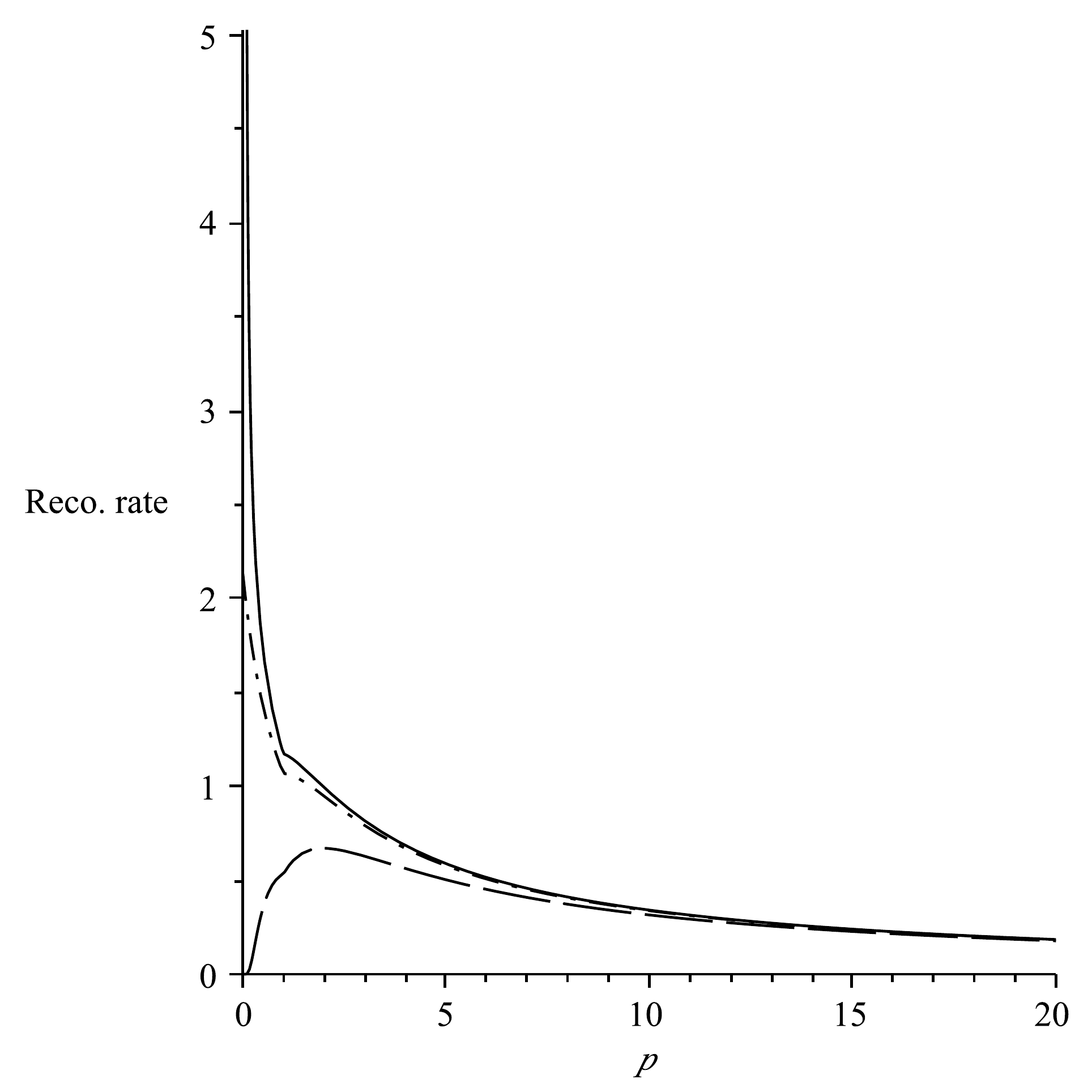}}
	\subfigure[]{\label{rr-b}   \includegraphics[width=7.5 cm]{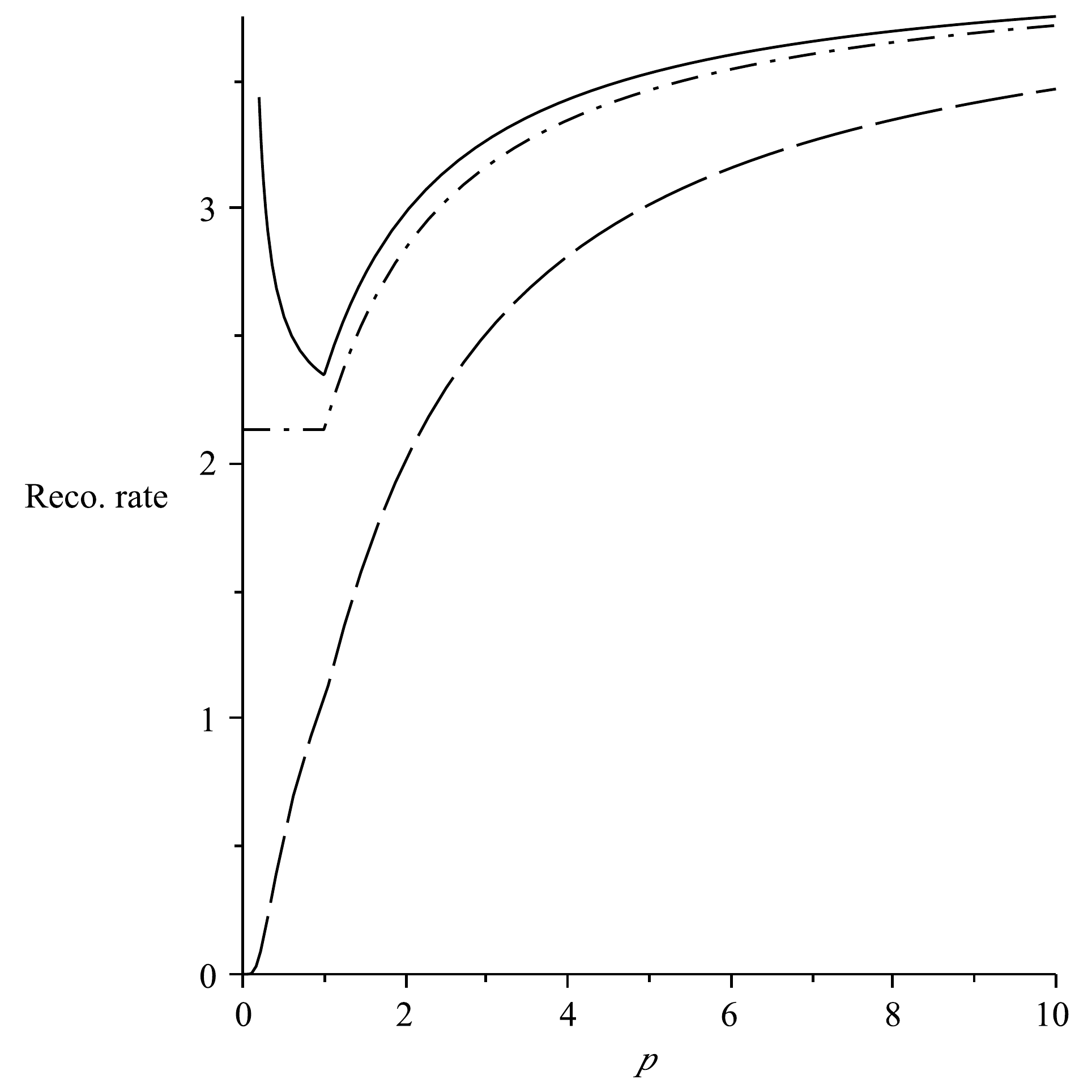} }
	\caption{\footnotesize Dependence of the reconnection rate on $p$, where the solid curve at {$a=1.5$}, dash-dotted curve at $a=1$, long dashed at {$a=0.5$}, for   (a) ${j=2j_0}$, (b) $j=j_0 (p+1)$, and parameters $\eta_{0}=\mu_0=B_0=j_0=L=1$.    }
    	\label{ali7}
\end{figure}

From equation (\ref{a}), we have 
\begin {eqnarray}
F&=&\int_{-a^{1/p}}^{a^{1/p}} E_{x} dx \nonumber \\ 
&=& \phi_1 \left\{
        \begin{array}{lc}
 \displaystyle\frac{16}{15} a^{1/p}, &  ~~0<p< 1,\\       
2a\left( 1+\displaystyle\frac{a^{4(p-1)}}{(4p+1)} -\frac{2a^{2(p-1)}}{(2p+1)} \right), &  ~~p\ge 1.         
        \end{array}\right.
        \label{kinrecrate}
        \end{eqnarray} 
 where $\phi_1$ is given by Equation (\ref{phi1}) as before. Note as a point of verification that this reduces to the expression found by \cite{fan2004} when $p=1$.
Here we consider the dependence of the reconnection rate on the parameter $p$ in two distinct cases---see Fig.~\ref{ali7}. First we set the parameter $j$ to be a constant, ${j=2j_0}$ say, which results in a current which is dependent on $p$. We then go on to consider the case where we set $j=j_0 (p+1)$, so that the current (${\bf J}=2B_0 j/(L(p+1))~\hat{{\bf x}}$) is independent of $p$. 

We will also consider the effect, in each of these cases, of taking different values for the parameter $a$, which controls the dimensions of the diffusion region. When $a=1$, the diffusion region is symmetric for all $p$, having circular cross-section in any plane of constant $z$. However, as stated above, the boundary of the diffusion region intersects each of the 3 coordinate axes at
\begin{equation}\label{bdaries}
\left\{
\begin{array}{cc}
x=\pm a^{1/p}, ~~ y=\pm a, ~~z=\pm b^{(p+1)/2}, &~0<p< 1,\\
x=\pm a^{1/p}, ~~ y=\pm a, ~~z=\pm b^{(p+1)/2p}, &~p\geq 1.
\end{array}\right.
\end{equation}
 Thus the diffusion region becomes asymmetric  in the $xy$-plane when $a\neq 1$ and $p \neq 1$. It will be seen later that this property is advantageous when comparing with the results of a numerical simulation.

\subsubsection{Reconnection rate as $p \rightarrow \infty$} 
In the limit $p\rightarrow\infty$, we observe from (\ref{bdaries}) that the diffusion region becomes approximately symmetric (exactly symmetric if $a=1$). The following all holds for all values of $a$. For the two choices of dependence for our parameter $j$ stated above, evaluating Eq.~(\ref{kinrecrate}) we find
\begin{equation}
\left. \lim_{p\to \infty} F ~ \right]_{{j=2j_0}} = 0
\end{equation}
\begin{equation}
\left. \lim_{p\rightarrow \infty} F ~ \right]_{j=j_0 (p+1)} = \frac{4j_{0}B_{0}\eta_{0}}{L \mu_{0}}
\end{equation}
Consider first the case where the parameter $j$ is chosen to be a constant, ${j=2j_0}$ (so that the current ${\bf{J}}={4B_{0}j_0}/{(L(p+1)\mu_{0})} \, \hat{\bf x}$). The magnetic field in this case is ${\bf B} \to (2B_0/L) (0, y, -z)$ as $p\to \infty$. That is, the magnetic field approaches a 2D X-point structure with zero current, and so the result above ($F \to 0$) is as expected.

By contrast, the reconnection rate approaches a constant finite value as $p\to \infty$ when $j=j_0 (p+1)$ (so that ${\bf{J}}={2B_{0}j_0}/{L\mu_{0}} \, \hat{\bf x}$). In this case the magnetic field is ${\bf B} \to (2B_0/L) (0, y-j_0 z, -z)$. So the configuration is that of a 2D X-point with a uniform current (proportional to $j_0$). As the diffusion region has only a finite extent along the direction of the current (${\hat {\bf x}}$), the reconnection rate is finite. Note that as expected it is proportional to the parameters $\eta_{0}, B_{0}/L$ and $j_{0}$, where $\eta_0$ is the resistivity at the null, and $2B_0 j_0/L\mu_0$ is the current modulus.  


\subsubsection{Reconnection rate as $p \rightarrow 0$}
We now turn to the opposite limit; $p\to 0$. Note that our two parameter choices ${j=2j_0}$ and $j=j_0(p+1)$ clearly reduce to the same situation (with $j_0$ replaced by $2j_0$) as the limit is approached. Setting $p=0$ the magnetic field is ${\bf B} = (2B_0/L) (x, -j_0 z, -z)$. We note that this field contains a neutral line in 3D (along $y=0$) which is anti-parallel to the direction of current flow---not a configuration associated with 2D reconnection. In fact the limit of Eq.~(\ref{kinrecrate}) is not well defined for all choices of our parameters. Therefore we consider that $p=0$ is not a physically relevant parameter choice and consider only the limit $p\to 0$.

As $p\to 0$, the magnetic field in the fan plane parallel to the current vector becomes strong, while the ${\hat {\bf y}}$-component becomes weak. Correspondingly, the flow across the fan surface becomes isolated to a small region near the fan, and weakens, see Fig.~\ref{ali2}. Furthermore,  in this case the diffusion region $D$ is highly anti-symmetric.

 Evaluating Eq.~(\ref{kinrecrate}) we find
\begin{equation}
 \displaystyle\lim_{p\to 0} F = \left\{
\begin{array}{cc}
 0, &a<1\\
 \frac{64B_0j_0\eta_0}{15L\mu_0},~~& a=1\\
 \infty, & a>1
\end{array}\right.
\end{equation}
(taking $j=2j_0$). For ${a<1}$ the extent of $D$ along the $x$-axis (direction of current flow) shrinks to zero. 
The result of the weak flow across the fan for small $p$ is therefore that the reconnection rate also approaches zero when $p \to 0$.  By contrast, for ${a>1}$  the boundaries of $D$ stretch to infinity along $x$ (see Eq.~(\ref{bdaries})). Correspondingly,  for $a>1$ the reconnection rate $F\to \infty$. Although the flow is still very weak across the fan, the diffusion region now has much larger extent in the $x$-direction, and so although the  flux reconnected per unit time per unit length in that direction decreases, the total flux reconnected increases.  When $a=1$, $D$ is symmetric, because the boundary is at $x=\pm a$, and the reconnection rate approaches a constant value as $p \to 0$.

The results discussed above show that depending on our choice of parameters there are various different ways in which the reconnection rate may depend on the asymmetry of the field ($p$). We now go on to perform simulations in the resistive MHD regime, in order to investigate which of these dependencies is relevant in a dynamically evolving plasma.

\section{Resistive MHD simulations} \label{numerical}
\subsection{ Computational setup   }
We now proceed to test the results of the mathematical model presented in the previous section by performing numerical simulations which solve the full set of resistive MHD equations.
We solve the MHD equations in the following form 
\begin{eqnarray} 
      \bf{E}&=&-\bf{v\times B +\eta J}\\
      \bf{J}&=&\nabla \times \bf{B}\\
       \frac{\partial \bf{B}}{\partial t}&=&- \nabla \times \bf{E}\\
        \frac{\partial \rho}{\partial t}&=&-\nabla \cdot(\rho \bf{v})\\
          \frac{\partial }{\partial t}(\rho {\bf{v}})&=&-\nabla \cdot(\rho \rho {\bf{v}}+\tau)-\nabla P+\bf{J \times B}\\
          \frac{\partial e}{\partial t}&=&-\nabla \cdot(e {\bf{v}})- P\nabla \cdot {\bf{v}}+Q_{visc}+Q_{J},
       \end{eqnarray}
 where {\bf{v,B,E}},$\eta,{\bf{J}} ,\rho,\tau,P, e,Q_{visc},Q_{J} $ are the velocity, magnetic field, electric field, resistivity, electric current, density, viscous stress, pressure, internal energy, viscous dissipation and Joule dissipation, respectively.
Here we provide a brief explanation of the method used for the numerical simulations. We run  simulations that are  similar to those described by \cite{pontinbhat2007a}. For more details on the numerical method, see \cite{nordlund1997, pontingalsgaard2007}. All simulations use numerical resolution of $128^3$ grid cells, a uniform resistivity model, and a so-called `hyper viscosity' model. This is calculated using a combined 2nd- and 4th-order method, which effectively `switches on' the viscosity only where grid-scale features develop in ${\bf v}$, in order to maintain code stability. In this way the effect of viscosity is minimised, and we focus on the effect of the resistivity (see \cite{nordlund1997}).

We consider an isolated three dimensional null point within our computational volume, which is driven from the boundary. We begin initially with a potential magnetic field
 \begin{equation} 
 {\bf{B}}=\frac{B_{0}}{L}\frac{2}{p+1}(x,py,-(p+1)z) \label{za}
 \end{equation} 
taking $B_{0}$=L=1 and  $\eta$=0.0007, constant, throughout. The computational domain has dimensions $[x,y,z]=[-3\dots 3, -3\dots 3, - 0.5\dots 0.5]$, with the magnetic field being line tied on all boundaries. In the kinematic model the spine and fan plane {are} not orthogonal, but in the our simulation at the outset they are orthogonal, which means the plasma is in  equilibrium. At $t=0$, the spine of the null point  lies  {in} $z$-direction, and the fan plane in the $z=0$ plane. A driving velocity is then assumed on the $z$-boundaries, which  {advects} the spine footpoints in opposite directions on the opposite boundaries. The spine is driven until the resulting disturbance reaches the null, resulting in {the} formation  {of} a current sheet as the magnetic field becomes stressed and distorted. 
The resulting configuration shares key properties with the configuration considered in the kinematic model: the spine and fan have collapsed toward one another generating a current parallel to the fan surface, and furthermore a localised non-ideal region is present around the null. After some time the driving velocity is reduced back to zero.
The explicit form taken for the driving velocity is defined by the streamfunction
\begin{equation}\label{incompprof}
\psi(z=\pm 0.5) = \pm 0.01 \left( \left(\frac{t-1.8}{1.8} \right)^4 -1 \right)^2  \sin \left( \frac{\pi x}{3} \right) \cos^2\left( \frac{\pi y}{6} \right) 
 e^{-8.9 (x^2+y^2)},
\end{equation}
$0 \leq t \leq 3.6$
(for more details, see \cite{pontinbhat2007a}). Below we compare  the kinematic solution and our simulation results.

\subsection{Current sheet} 
\begin{figure}[t]
\begin{center}
\subfigure[]{\label{bstruc_tzero}\includegraphics[width=4.5cm]{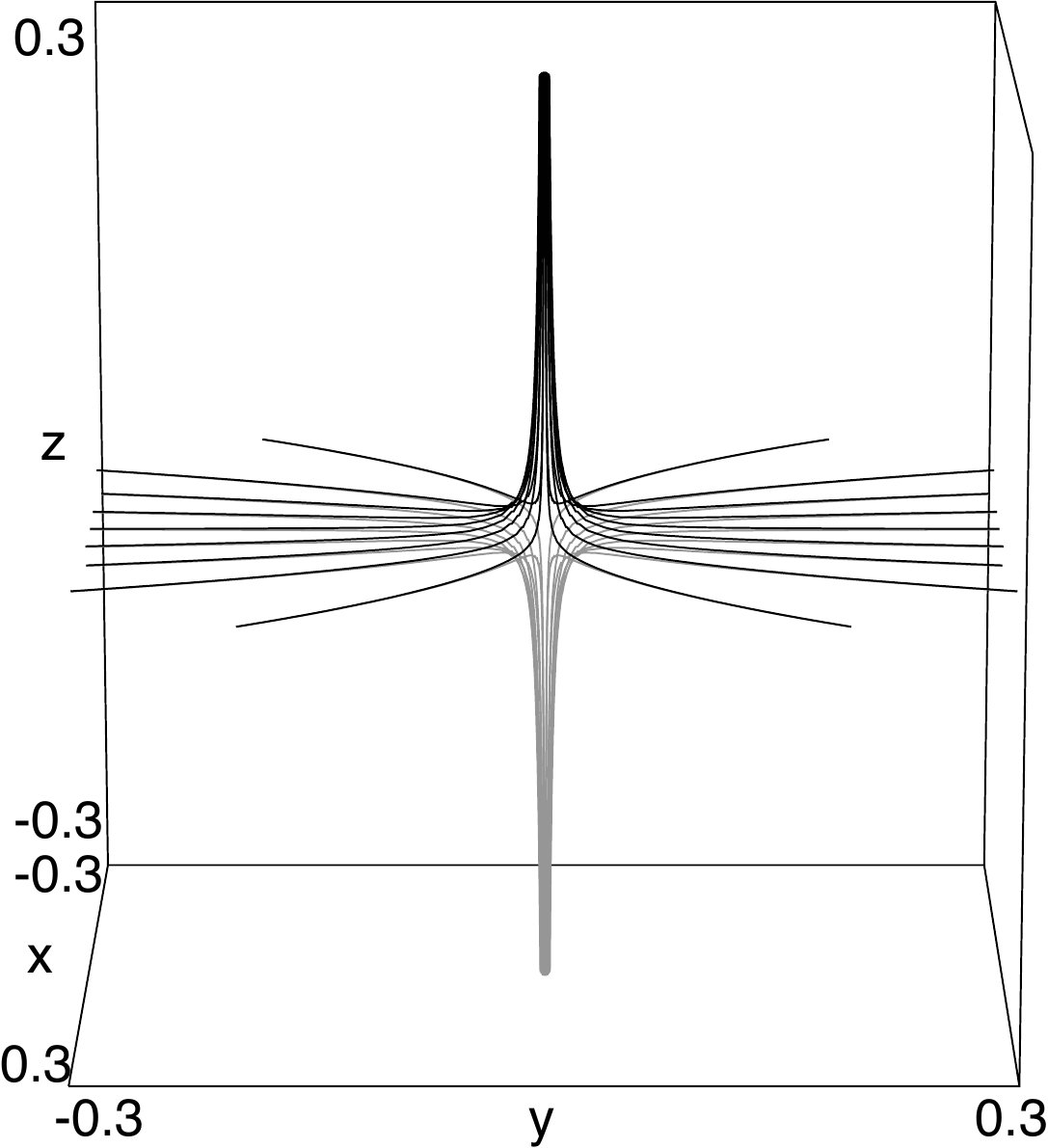}}
\subfigure[]{\label{bstruc_sheet}\includegraphics[width=4.5cm]{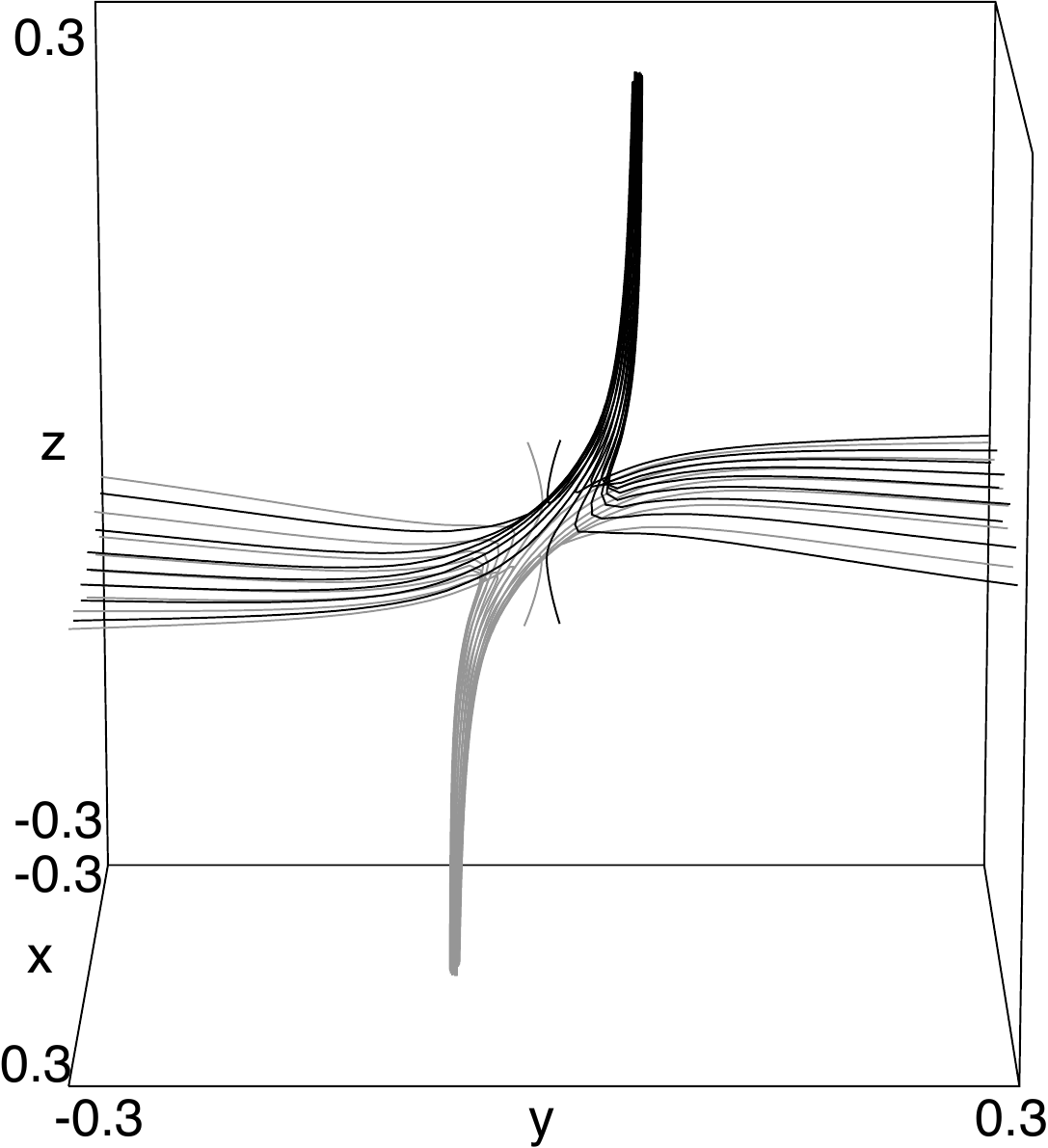}}
\subfigure[]{\label{j_xy}\includegraphics[width=4.3cm]{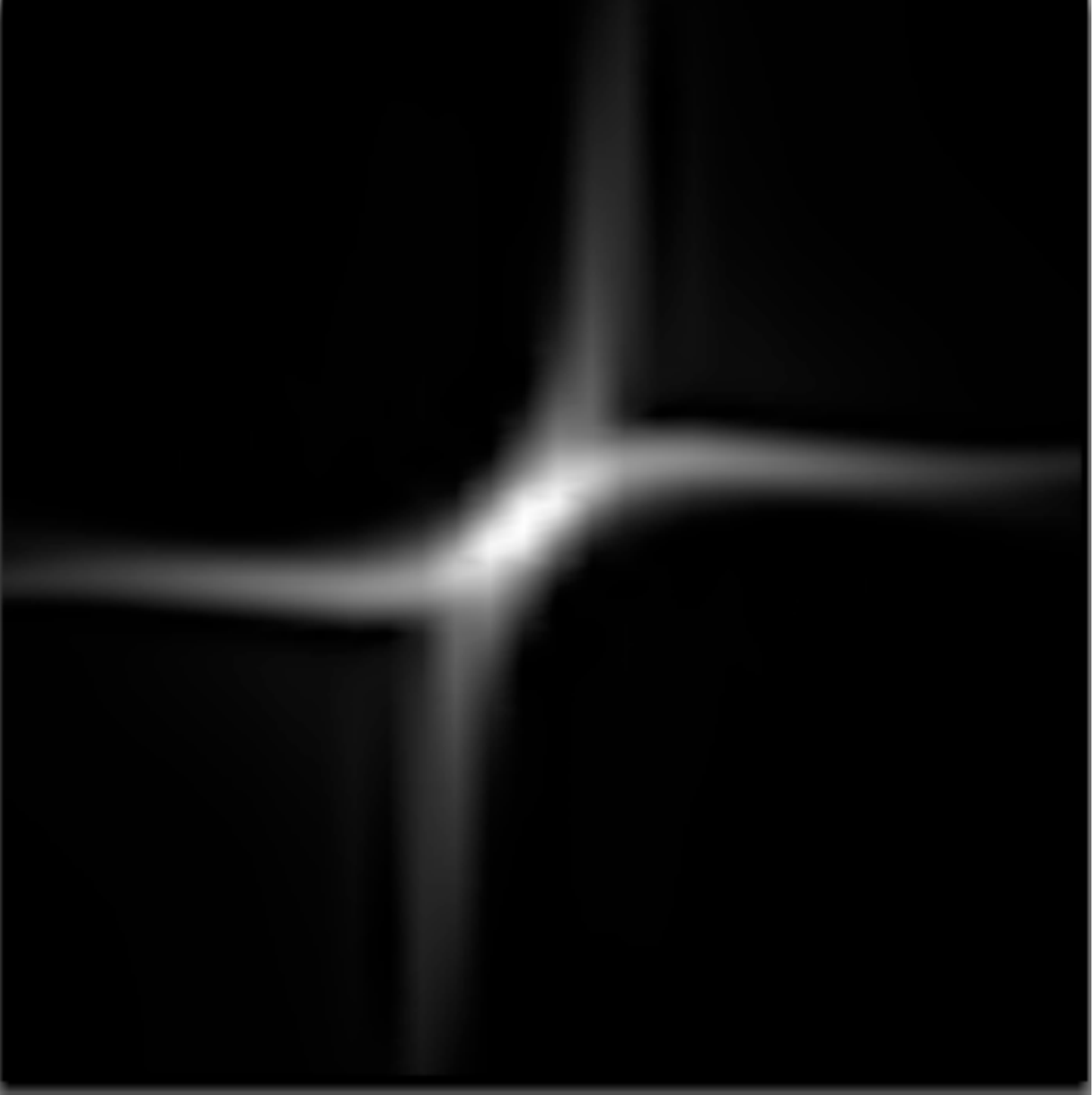}}
\end{center}
\caption{\footnotesize Structure of the magnetic field for the run with $p=2$ (a) at $t=0$, and (b) at the time of maximum current ($t=3.0$), once the magnetic field has locally collapsed to form a current sheet. The black field lines are traced from around the spine for $z>0$, while the grey field lines are traced from $z<0$. (c) Grayscale showing $|{\bf J}|$ in the $x=0$ plane at the same time as (b).}
\label{jsheetpics}
\end{figure}

In order to simplify the discussion  we will initially explain the behaviour of  the current at one value of $p$ ($p=2$). We first examine the temporal evolution of current in the volume. In the  beginning the spine and fan are orthogonal, but then the angle between them  begins to change, reaching a minimum value once the current sheet forms. In other words the null collapses from a perpendicular $X$- type null point, with the angle between the $X$ becoming greatly reduced, see Fig.~\ref{jsheetpics}. After the boundary driving ceases the current begins to decrease again, and the spine and fan relax back towards their initial perpendicular state, see Fig.~\ref{ali8}.  
  
\begin{figure}[t]
\begin{center}
\subfigure[]{\label{fig:edge-3a}\includegraphics[width=0.24\textwidth]{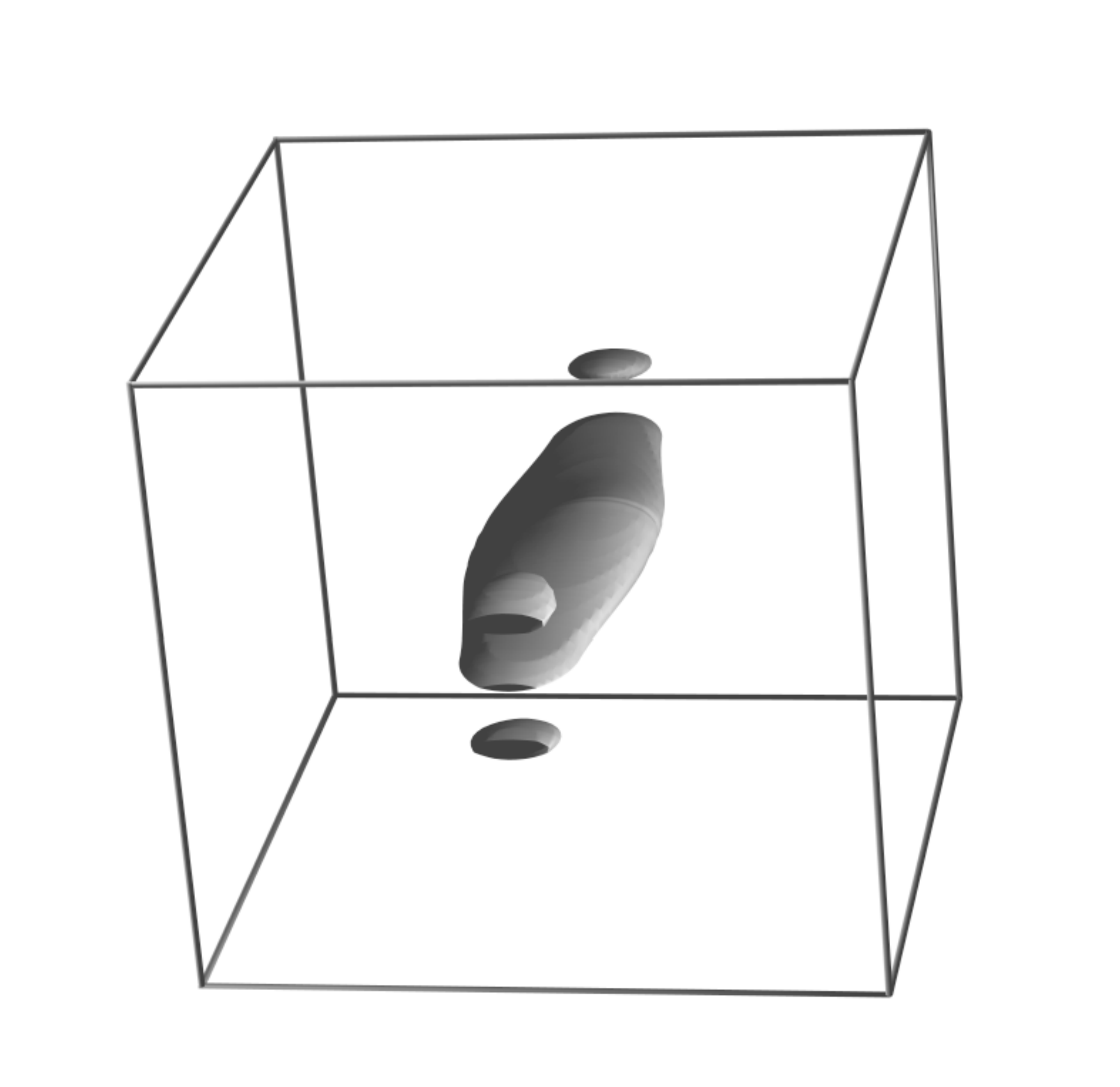}}
\subfigure[ ]{\label{fig:edge-3b}\includegraphics[width=0.24\textwidth]{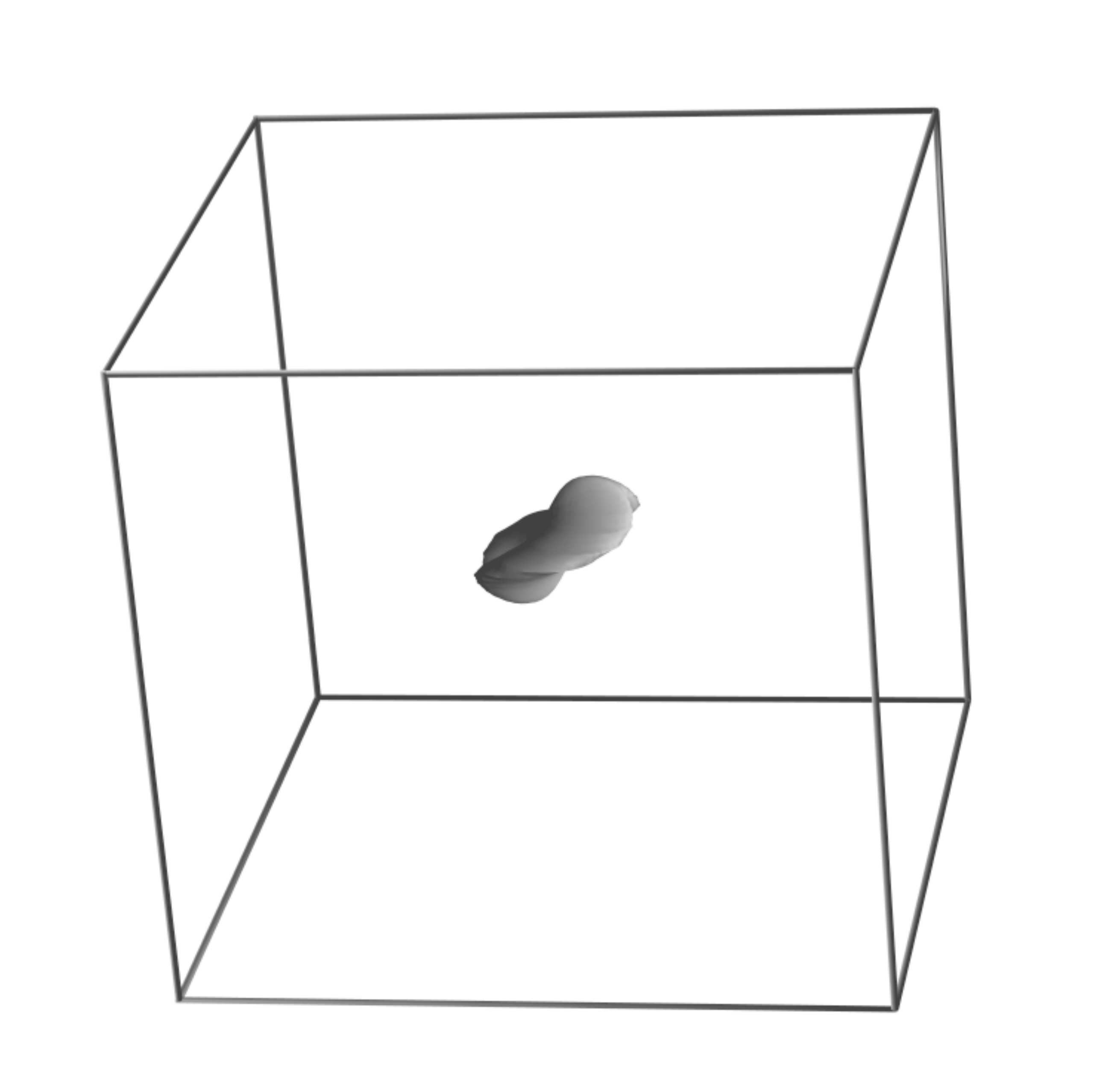}}
\subfigure[]{\label{fig:edge-3c}\includegraphics[width=0.24\textwidth]{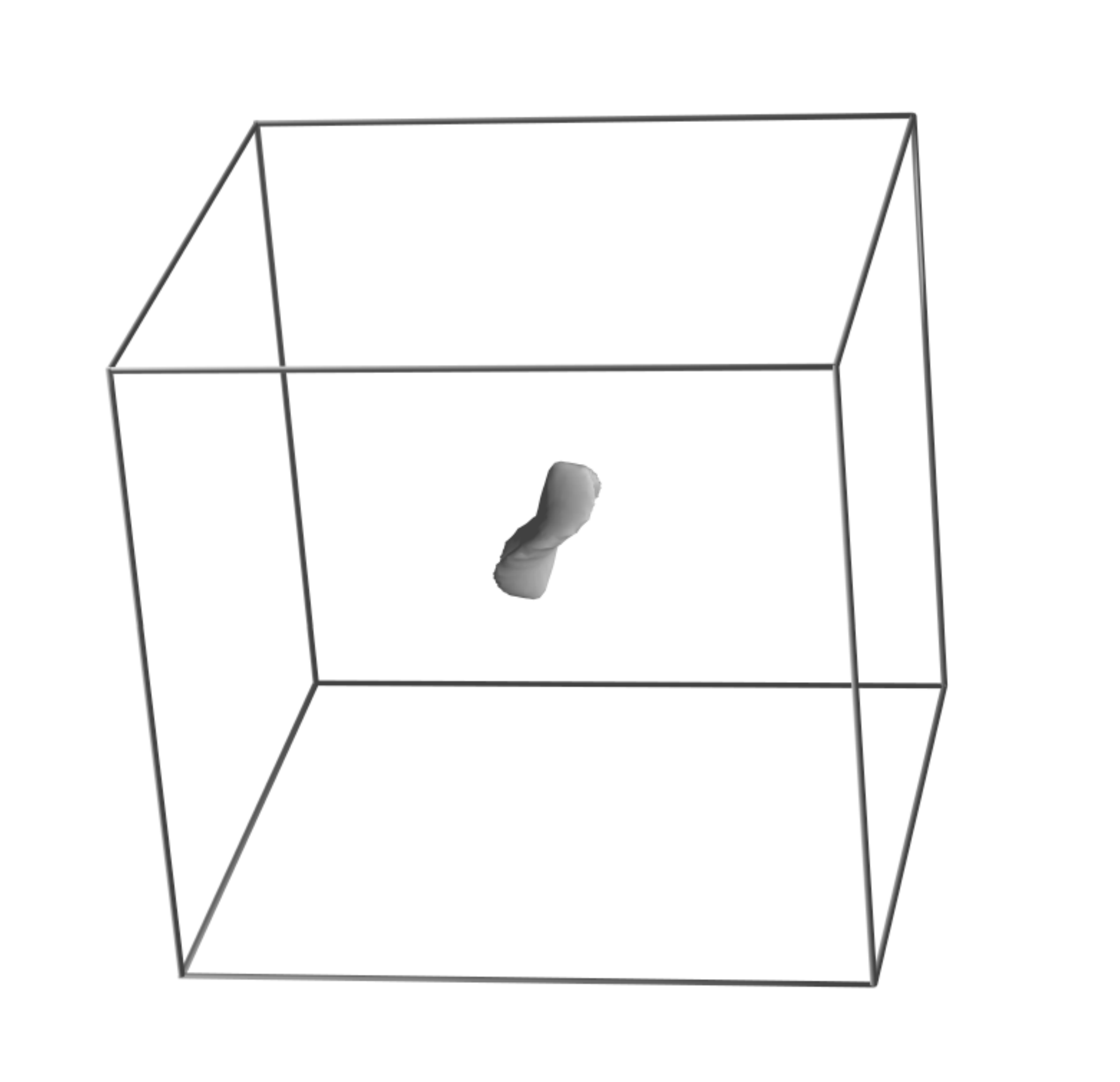}}
\subfigure[]{\label{fig:edge-3d}\includegraphics[width=0.24\textwidth]{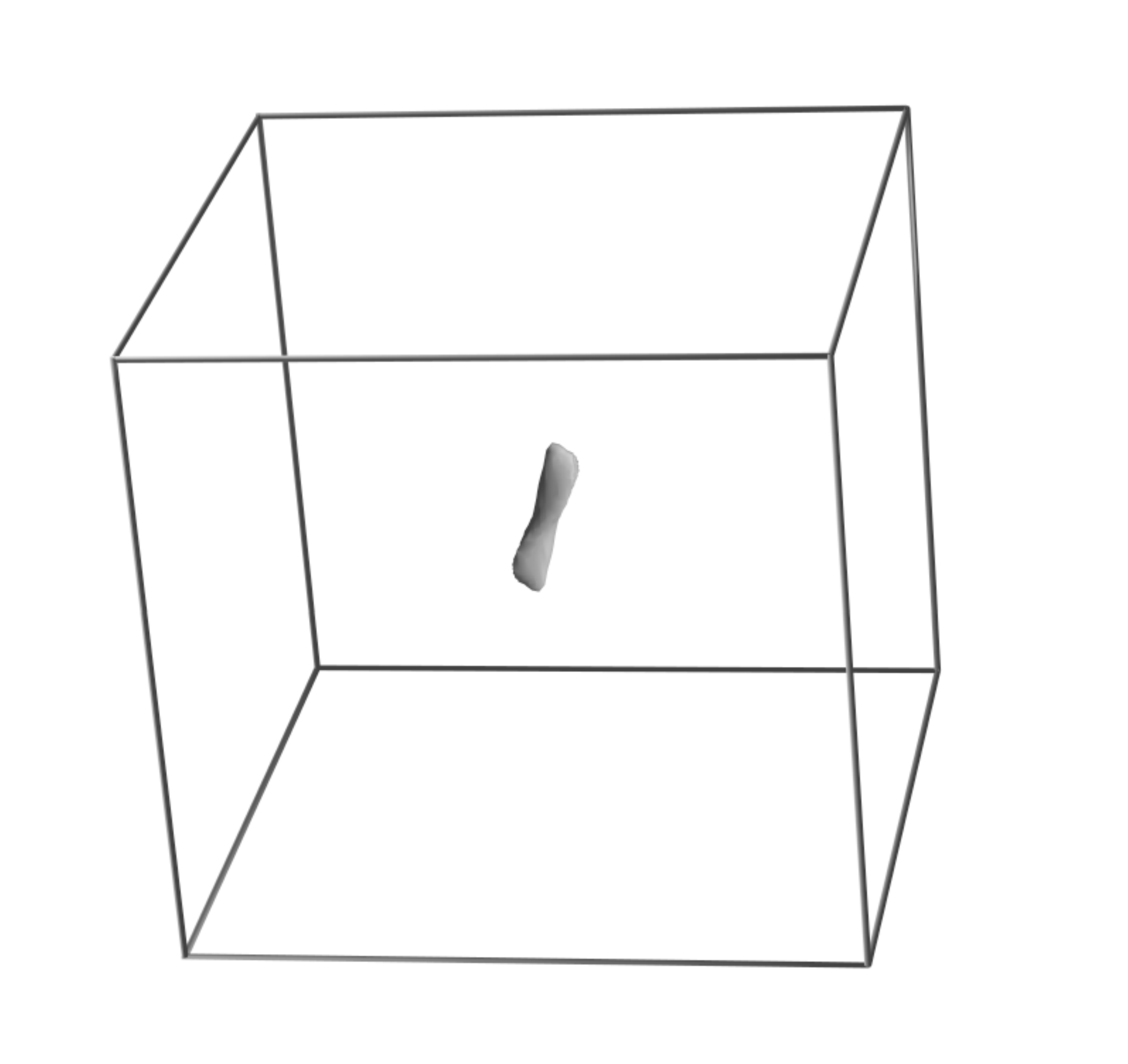}}
\subfigure[]{\label{fig:edge-3e}\includegraphics[width=0.3\textwidth]{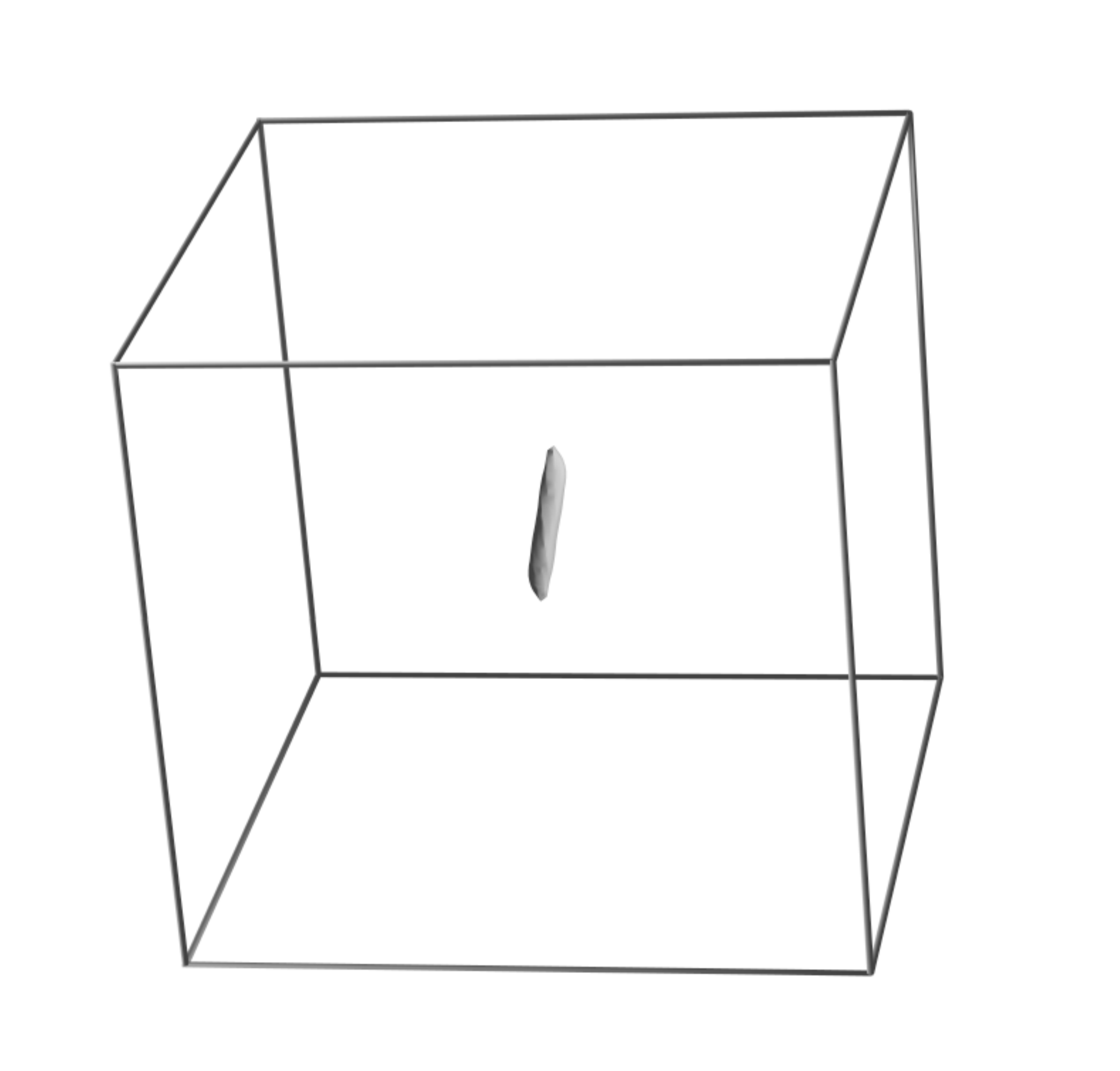}}
\end{center}
\caption{\footnotesize Isosurfaces of $|{\bf J}|$, at 50\% of the maximum at that time, showing dimensions of the current sheet for different $p$ at the time of maximum $|{\bf J}|$: (a)  $p=0.1$ ($t=3.8$), (b)  $p=0.5$ ($t=3.5$), (c) $p=1$ ($t=3.3$), (d) $p=2$ ($t=3.0$) and (e) $p=10$ ($t=3.2$). Plot dimensions are $[x,y,z]=[\pm 2.7, \pm 1, \pm 0.17]$.}
\label{ali10}
\end{figure}

We now discuss how the current sheet formation, as described above, depends on the value of $p$. Fig.~\ref{ali10} illustrates the dimensions of the current sheet for various values of $p$ at the time when the current modulus is of maximum value.  For the case investigated previously, $p=1$, the sheet was found to be approximately of equal dimensions along $x$ and $y$, the two coordinate directions associated with the fan surface. Looking at  {figure \ref{ali10} we see} a large difference between the geometry of the current sheet at $p=0.1$ and at $p=10$ at the maximum current. We find the current sheet at $p=0.1$ is large, being very extended along the  $x$-axis, that is, the direction along which ${\bf J}$ and the parallel electric field lie. However, this length decreases when  $p$ is  increased.  The results suggest that when the value of $p$ approaches zero, the current sheet will grow indefinitely in the plane perpendicular to the shear, i.e.~the direction of current flow through the null ($x$-direction). Note that with respect to the field strength in the fan plane, decreasing $p$ corresponds to weaker magnetic field strength along the $x$-direction. Thus the extension of the current sheet could be attributed to the fact that the weak field region extends in that direction and the  magnetic field becomes less able to resist the collapse to form the current layer. That is, when the magnetic field parallel to the current becomes weaker there is less magnetic pressure associated with this `guide field' component in the current sheet away from the null, and the current sheet is able to extend further away from the null.
    
\subsection{Maximum current attained}  
\begin{figure}[t]
\begin{center}
\subfigure[]{\label{ali8}\includegraphics[width=0.45\textwidth]{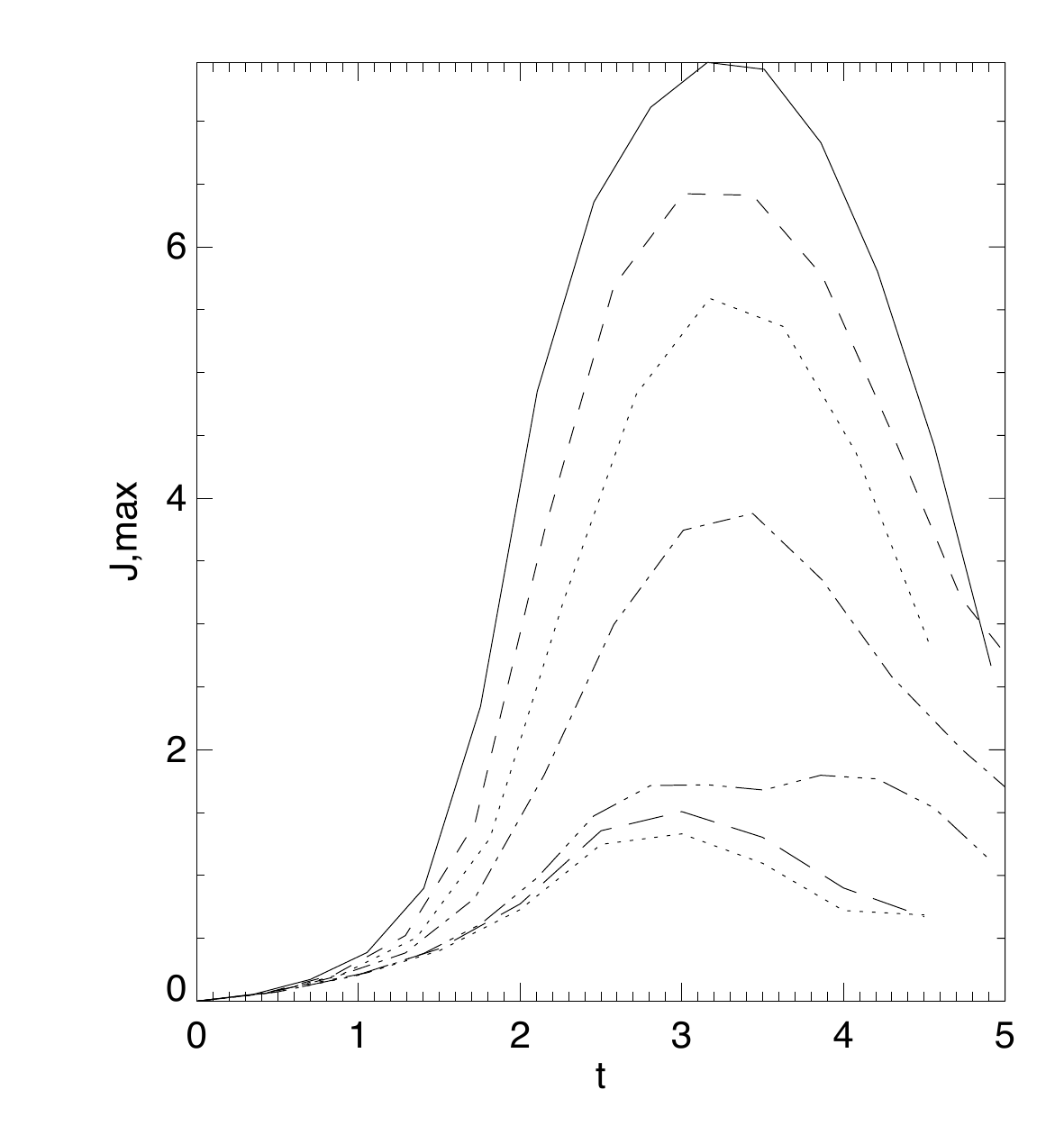}}
\subfigure[ ]{\label{ali9}\includegraphics[width=0.45\textwidth]{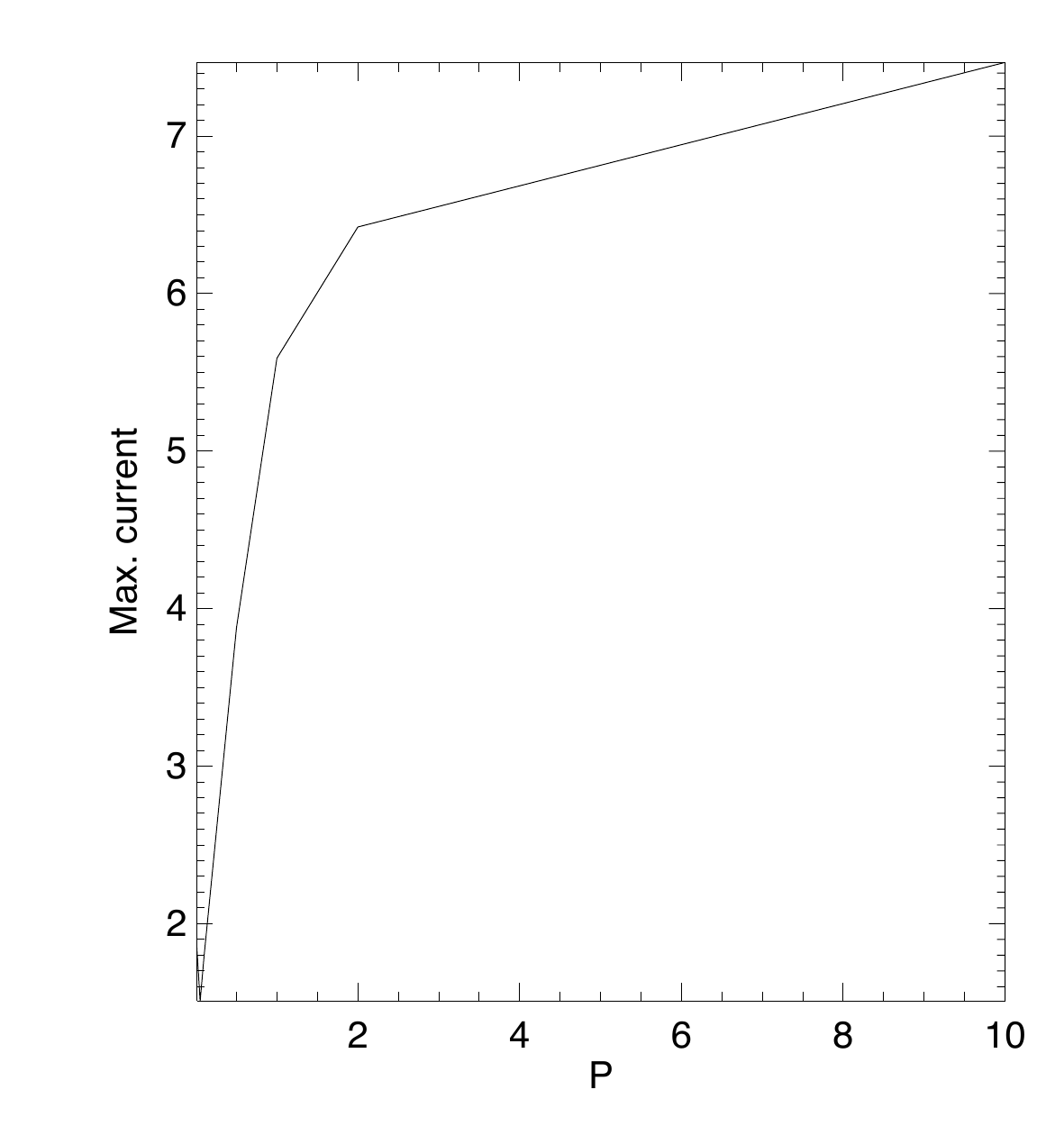}}
\end{center}
\caption{\footnotesize  (a) Evolution of the maximum value of $|{\bf J}|$  in time with different $p$; $p=0.01$ (dotted),  $p=0.05$ (long dashed), $p=0.1$ (dash dot dot), $p=0.5$ (dash dot), $p=1$ (dotted), $p=2$ (dashed) and $p=10$ (solid). (b) Peak spatial and temporal value of $|{\bf J}|$ for different $p$.   }
\label{ali14}
\end{figure}
 Figure \ref{ali8} illustrates the evolution of the current modulus maximum within the domain in time, for runs with different values of the parameter $p$. We notice in each case that the peak current grows sharply in time to reach a maximum value, and then decreases again, as discussed above. Figure \ref{ali9} shows  the maximum value in time of current plotted against $p$. We observe that there is a positive correlation between $p$ and the maximum current -- in other words, when $p$ increases then the maximum value of the current that is attained also increases. Furthermore there is  a negative correlation between the size of the current sheet and the value of $p$, see  figure (\ref{ali10}). Thus, although the current becomes more localised as $p$ increases, it also becomes more intense. 
One important point to note is that of course the effective value of $p$ will change during the simulations as the magnetic field is deformed. This can be confirmed be calculating the eigenvalues of $\nabla {\bf B}$ at the null as the simulations proceed. We find that the relative change is small -- of order 1\% for $p=0.1$ and order 10\% for $p=10$. Thus the ordering of the values of $p$ that we selected for our simulations is preserved and the trends that are observed for the $p$-dependence will be unaffected.

 \subsection{Reconnection rate}
\begin{figure} 
   	\centering
	\subfigure[]{\includegraphics[width=5.5 cm]{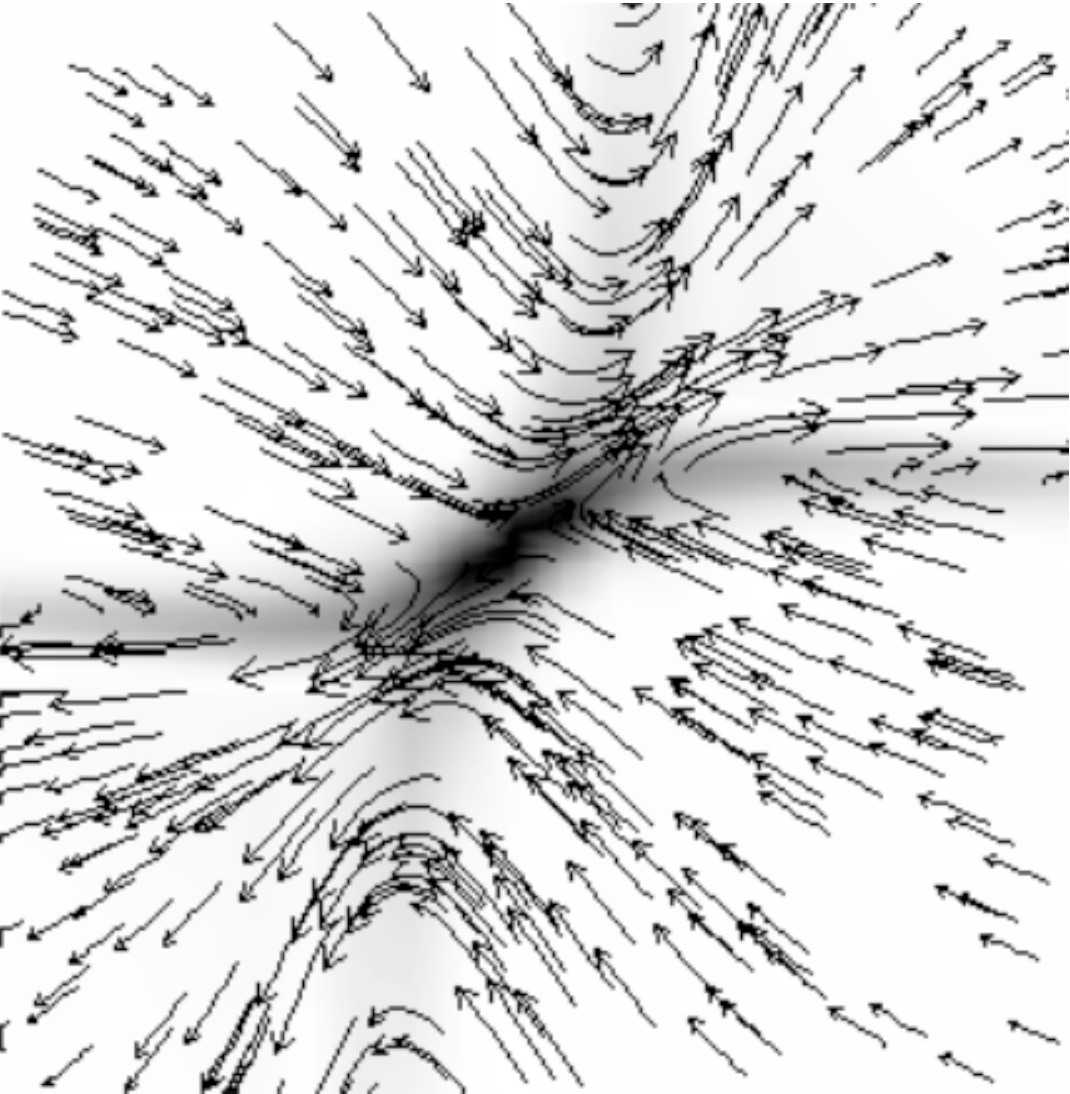}}
	\subfigure[]{\includegraphics[width=5.5 cm]{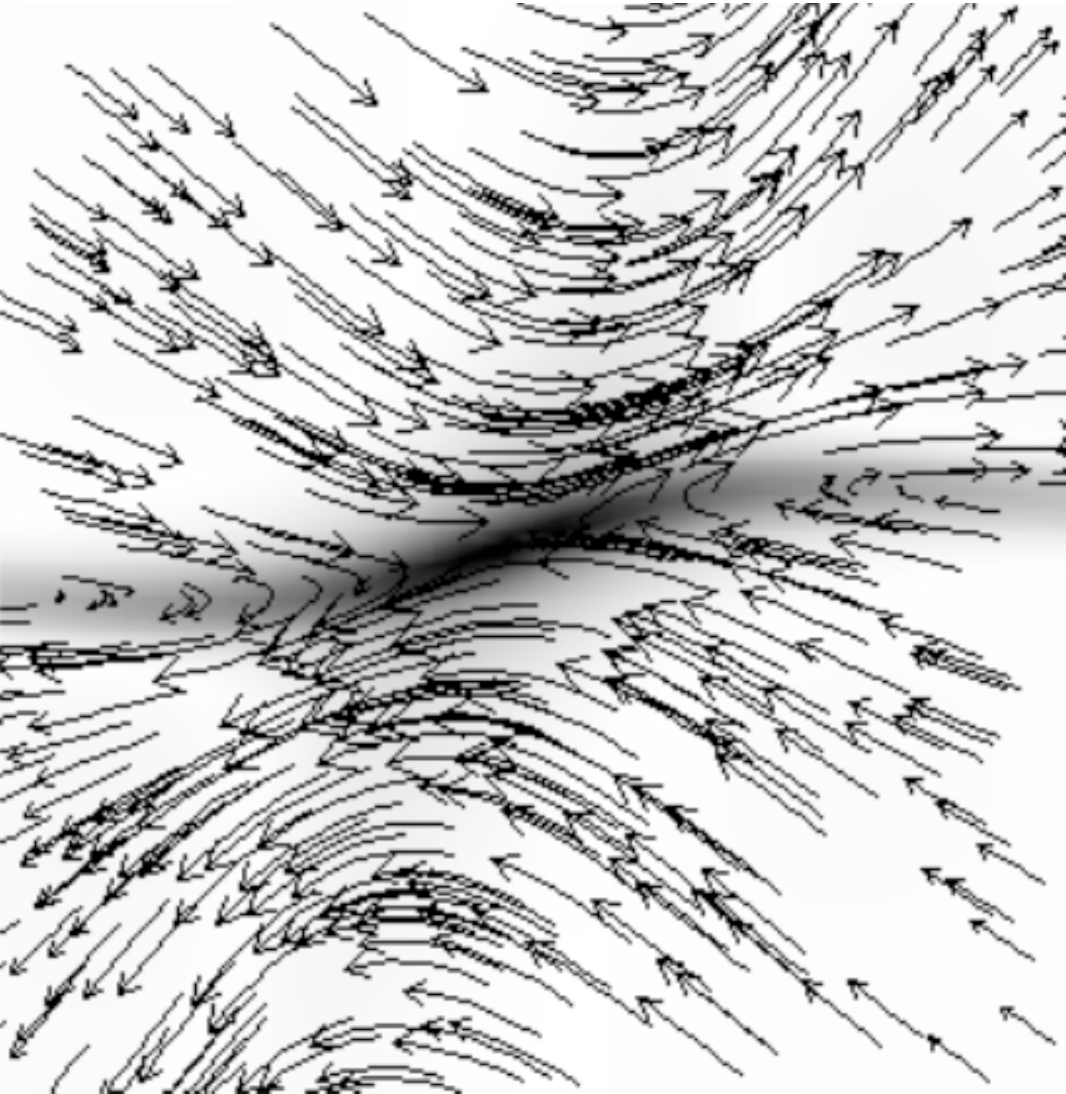} }
	\caption{\footnotesize Plasma velocity in the $x=0$ plane for $[x,y]=[\pm0.25, \pm0.25]$ and (a) $p=2$ (b) $p=0.5$. Background shading shows the current density.}
    	\label{vnum}
\end{figure}

The nature of the plasma flow, and the resulting qualitative structure of the reconnection process, are found to be  independent of the value of $p$. Specifically, we find plasma flow across both the spine line and fan plane of the null for all values of $p$. Figure \ref{vnum} shows the plasma flow for two different values of $p$. Comparing with Figure \ref{ali2}, we see that the trend for the geometry of the flow is the same as in the kinematic solution. Specifically, for large $p$, the flow exhibits a relatively symmetric stagnation structure (in the $x=0$ plane). For smaller $p$ the flow across the fan becomes confined to a narrower region, and comparatively weaker with respect to the flow across the spine.
 
In this section we calculate the reconnection rate, i.e.~the amount of flux transported across the fan surface, 
 as before by integrating the electric field component parallel to the magnetic magnetic field ($E_{\parallel}$). Similarly to above, by symmetry, the integral is performed along the field line lying along the $x$-axis,
where since we are in the resistive MHD regime we have $ E_{\parallel}={\eta \bf{J\cdot B}}/{|{\bf{B}}|}$.

In figure \ref{ali14} we show the evolution of the reconnection rate in time for different values of $p$. Initially the rate clearly stay constant (zero) in time, i.e.~during the early evolution, between $t=0$ and $t=1.$  Later, it starts to develop until it gains its maximum value, and then begins to decrease. This follows the same pattern as the evolution of the current, being indicative of the fact that the null point  collapses  to form the current sheet and reconnection occurs, and then the system relaxes once the driving ceases. 
It is clear from Fig.~\ref{ali14} that the maximum reconnection rate attained increases as the value of $p$ is decreased.  It is worth emphasising here that although our intuition tells us that there is positive correlation between current and reconnection rate, by contrast in this study we notice the inverse is true, i.e.~when the peak current increases the reconnection rate decreases. This is because the diffusion region  stretches when $p$ tends to zero in the direction where the $E_{||}$ lies. Therefore  the rate increases even though the current  decreases, since the integrand in Eq.~(\ref{recrateeq}) is non-zero over a much larger portion of the $x$-axis.      

If we finally compare our results with those of the incompressible model of \cite{craig1998}, we find their results differ from ours in terms of the dependence of the peak current on $p$. In particular, they found that (in terms of our parameters) the maximum current decreases when $p$ goes to infinity.  This may be down to the very different geometries of the current sheet in the two models (the current sheet in their incompressible model is planar and extends to infinity in all directions along the fan for all values of $p$). However, it is of interest to note that we actually find the same dependence of reconnection rate on $p$, i.e.~as $p$ decreases the reconnection rate increases (since in fact we find a negative correlation between $J_{max}$ and the reconnection rate as $p$ is varied).
      
\begin{figure}
\begin{center}
\subfigure[]{\label{ali12}\includegraphics[width=0.45\textwidth]{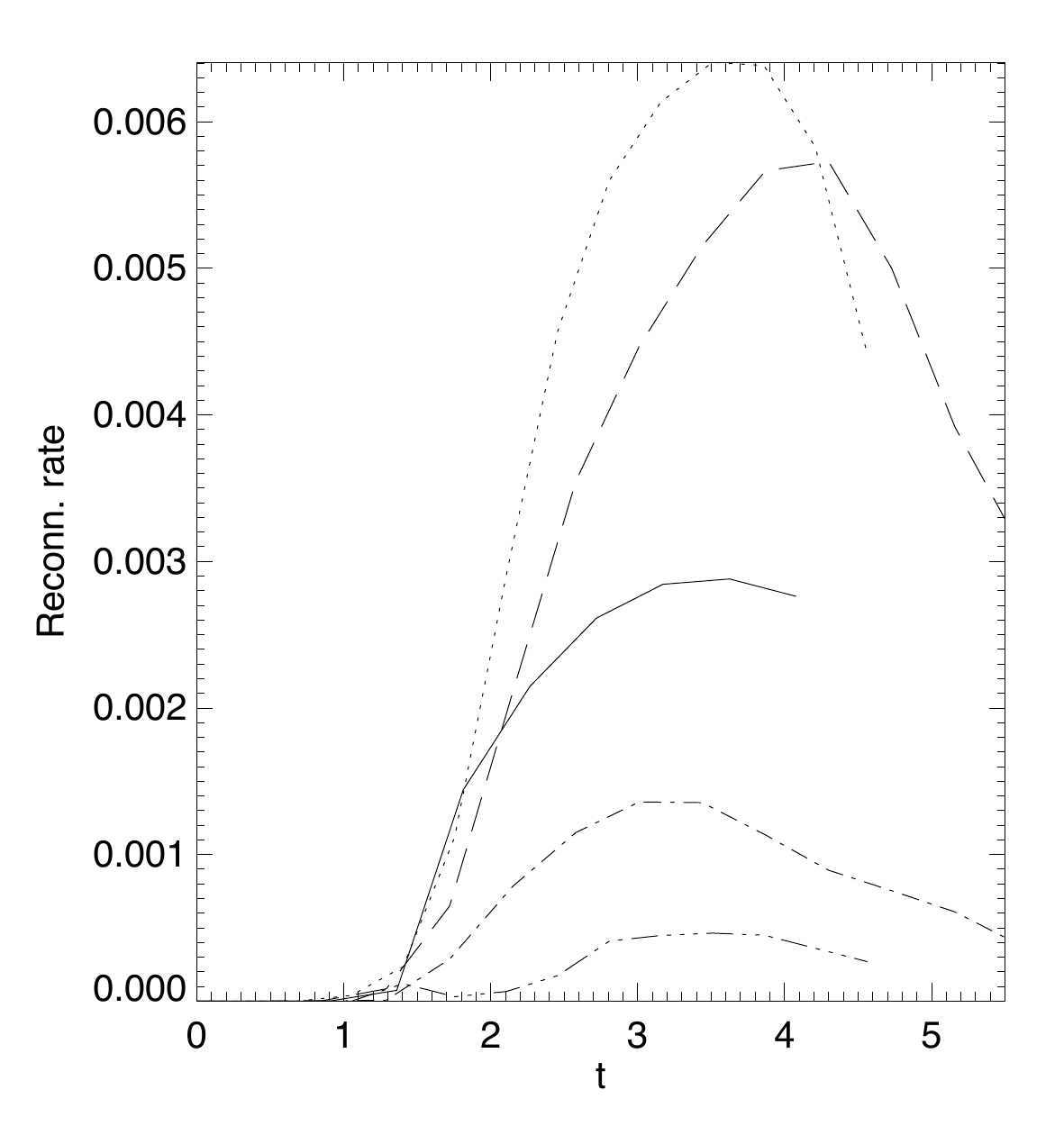}}
\subfigure[ ]{\label{ali13}\includegraphics[width=0.45\textwidth]{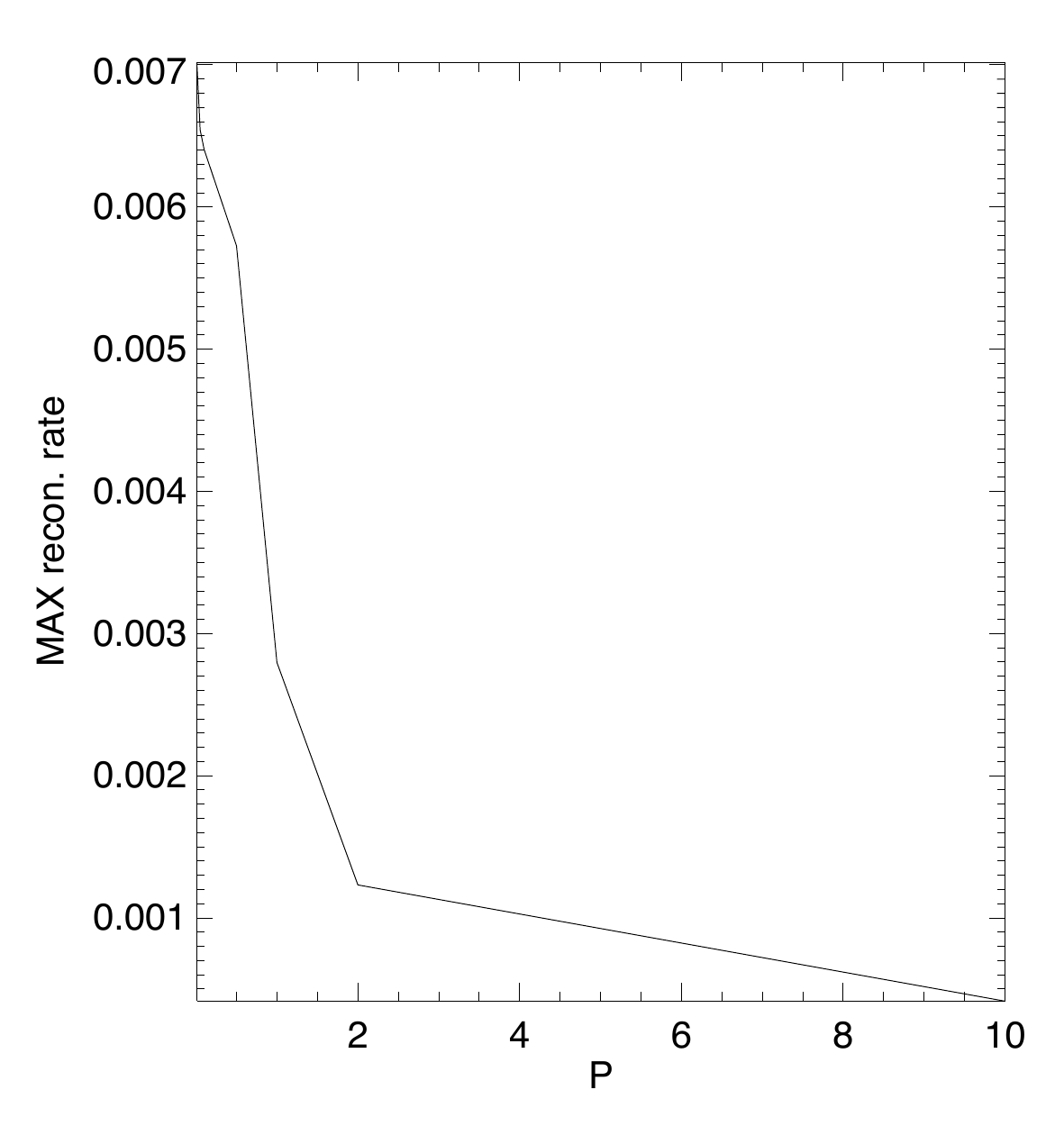}}
\end{center}
\caption{\footnotesize  (a) Reconnection rate  at different values of $p$, where the dotted curve is the reconnection rate  at $p=0.1$, the long dashed curve at $p=0.5$, the solid curve at $p=1$,  the dashed dot curve at $p=2$ and the dashed dot dot curve at $p=10$. (b) Variation of the maxiumum reconnection rate with the parameter $p$.}
\label{ali14}
\end{figure}

\section{Conclusions and Discussion}\label{conc}
In this paper we have investigated the effect of the symmetry of the magnetic field on magnetic reconnection at an isolated null point. We concentrate on the so-called spine-fan mode of 3D null point reconnection \citep{priest2009}. In a future paper we will go on to consider the `torsional spine' and `torsional fan' modes, which involve a current flowing parallel to the spine of the null point.

In the first part of the paper we discussed a steady solution of a subset of the resistive MHD equations, where the magnetic null point was defined by {\bf{B}}=$B_{0}^{\prime}(x,py - jz,- (p+1)z)$. This magnetic field has current aligned to the fan surface of the null point, and \cite{fan2004} investigated this situation in the non-generic symmetric case $p=1$ (repeated eigenvalues). In this work we use $p$ as a parameter.  
By necessity, as the dynamics of the system are not included in this steady-state kinematic solution, a current is imposed, which has the same orientation at the null as found in the simulations (the orientation of ${\bf J}$ at the null has been shown to be the crucial quantity in determining the topological structure of the reconnection process \citep{spine2004, fan2004}). In order to have a localised diffusion region around the null, we artificially imposed a localised resistivity.
We found the nature of the plasma flow, and the resulting qualitative structure of the reconnection process, to be the same as found in the symmetric case. Specifically, we found plasma flow across both the spine line and fan plane of the null for all values of $p$. 

We then described the results of a computational resistive MHD simulation in which we investigated the nature of the MHD evolution for different values of the parameter $p$ (the ratio of the fan eigenvalues). Since in this case the full set of MHD quations was solved self-consistently, we began with an equilibrium potential magnetic null point (with ${\bf J}={\bf 0}$). The system was then driven away from this equilibrium in such a way as to induce a local collapse of the null leading to current sheet formation and spine-fan magnetic reonnection. The resulting configuration shares key properties with the analytical solution: the spine and fan are non-orthogonal with a current flowing parallel to the fan surface, and a localised diffusion region is focussed at the null.
Also, in both cases the flow in the $yz$-plane exhibits a stagnation-point structure. There is agreement between the model and the simulations, in that for large $p$ the  stagnation structure is relatively symmetric, while for smaller $p$ the flow across the fan becomes confined to a narrower region, and weaker compared with the flow across the spine.

One of the major results that arises from the sequence of simulations is that both the peak intensity and the dimensions of current sheet are strongly dependent on the symmetry/asymmetry of the field in the fan surface, or in other words on the value of $p$. In terms of the sheet dimensions, the length along the direction of current flow at the null increases when $p$  goes to zero, i.e.~the diffusion region is stretched in the $x$-direction when $p$  tends to zero. In the kinematic solution it was also possible by choosing the correct parameters to have the diffusion region dimensions have such a $p$-dependence. In order for our kinematic solution to be physically relevant, this implies that the parameter $a$ in our solution should be chosen such that $a>1$.  Furthermore, as there is little difference in the size of the diffusion region in $z$ for different $p$ in the simulations, we should take $b=1$ in our mathematical model.
      
In addition to the current sheet at the null, we examined the reconnection rate in both cases. In order to compare the results, in light of the discussion above, we consider the parameter regime $a>1$ in the kinematic solution. When $a>1$ the reconnection rate $\to \infty$ as $p\to 0$. On the other hand, as $p\to \infty$ the reconnection rate approaches either zero or a constant finite value, depending on whether the current falls to zero or remains fixed, respectively, as $p$ is increased (see Figure \ref{ali7}). Turning to the simulations, as shown in Figure \ref{ali14} the reconnection rate indeed becomes very large as $p\to 0$, in agreement with the kinematic model. In addition, as $p$ becomes large the current at the null falls, and the reconnection rate appears to asymptotically approach some small value, also in agreement with the kinematic model. Whether this value is finite or zero is not possible to tell within the restrictions of the present simulations. 

The results of both the mathematical model and simulations reveal that the symmetry/asymmetry of the magnetic field in the vicinity of a null can have a profound effect on the geometry of any associated reconnection region, and the rate at which the reconnection process proceeds.

\section{Acknowledgments}
We would like to thank G. Hornig, A. L. Wilmot-Smith and E.~R.~Priest for helpful and stimulating discussions. A.K.H. Al-Hachami was supported in this work  by a grant from the Iraqi Government. Computational simulations were developed in conjunction with K. Galsgaard, and run on the MHD Computing Consortium's Beowulf cluster. 
 \bibliographystyle{apalike}
  \bibliography{dr}
\end{document}